\documentclass[10pt,aps,prd,showpacs,eqsecnum,nofootinbib,%twocolumn,
amsmath,amssymb,floats]{revtex4}

\usepackage{dcolumn}
\usepackage{graphics}

\allowdisplaybreaks

\newcommand{\be}{\begin{equation}}
\newcommand{\bl}{\mbox{\boldmath$\ell$}}
\newcommand{\bse}{\begin{subequations}}
\newcommand{\bsigma}{\mbox{\boldmath$\sigma$}}

\newcommand{\cpipii}{{\mathbf{p}}_1\times{\mathbf{p}}_2}

\newcommand{\ee}{\end{equation}}
\newcommand{\ese}{\end{subequations}}

\newcommand{\hnr}{H^{\mathrm{NR}}}

\newcommand{\hsoi}{\hat{H}^\mathrm{so}_\mathrm{LO}}
\newcommand{\hsoii}{\hat{H}^\mathrm{so}_\mathrm{NLO}}
\newcommand{\hsoip}{H^{\prime\,\mathrm{so}}_\mathrm{LO}}
\newcommand{\hsoiip}{H^{\prime\,\mathrm{so}}_\mathrm{NLO}}
\newcommand{\hsoipp}{H^{\prime\prime\,\mathrm{so}}_\mathrm{LO}}
\newcommand{\hsoiipp}{H^{\prime\prime\,\mathrm{so}}_\mathrm{NLO}}

\newcommand{\ncpi}{{\mathbf{n}}_{12}\times{\mathbf{p}}_1}
\newcommand{\ncpii}{{\mathbf{n}}_{12}\times{\mathbf{p}}_2}
\newcommand{\np}{({\bf n}\cdot{\bf p})}
\newcommand{\npi}{(\mathbf{n}_{12}\cdot\mathbf{p}_1)}
\newcommand{\npii}{({\bf n}_{12}\cdot{\bf p}_2)}
\newcommand{\npv}{{\bf n}\times{\bf p}}

\newcommand{\pa}{\partial}
\newcommand{\piipii}{{\bf p}_2^2}

\newcommand{\pipi}{{\bf p}_1^2}
\newcommand{\pipii}{({\mathbf{p}}_1\cdot{\mathbf{p}}_2)}

\newcommand{\pp}{{\bf p}^2}

\newcommand{\snp}{\big(\bar{S},n,p\big)}
\newcommand{\ssnp}{\big(\bar{S}^{*},n,p\big)}

\begin{document}

\title{Effective one body approach to the dynamics of two spinning
black holes with~next-to-leading order spin-orbit coupling}

\author{Thibault Damour}
\email{damour@ihes.fr}
\affiliation{Institut des Hautes \'Etudes Scientifiques,
91440 Bures-sur-Yvette, France}

\author{Piotr Jaranowski}
\email{pio@alpha.uwb.edu.pl}
\affiliation{Faculty of Physics,
University of Bia{\l}ystok,
Lipowa 41, 15--424 Bia{\l}ystok, Poland}

\author{Gerhard Sch\"afer}
\email{gos@tpi.uni-jena.de}
\affiliation{Theoretisch-Physikalisches Institut,
Friedrich-Schiller-Universit\"at,
Max-Wien-Pl.\ 1, 07743 Jena, Germany}

\date{\today}

\begin{abstract}

Using a recent, novel Hamiltonian formulation
of the gravitational interaction of spinning binaries,
we extend the Effective One Body (EOB) description of the dynamics of
two spinning black holes to next-to-leading order (NLO) in the
spin-orbit interaction. The spin-dependent EOB Hamiltonian
is constructed from four main ingredients: (i) a transformation between the ``effective''
Hamiltonian and the ``real'' one, (ii) a generalized effective Hamilton-Jacobi equation
involving higher powers of the momenta, (iii) a Kerr-type effective metric (with Pad\'e-resummed
coefficients) which depends on the choice of some basic ``effective spin vector'' $\bf{S}_{\rm eff}$,
and which is deformed by comparable-mass effects,
and (iv) an additional effective spin-orbit interaction term involving another spin vector $\bsigma$.
As a first application of the new, NLO spin-dependent EOB Hamiltonian, we compute the binding energy
of circular orbits (for parallel spins) as a function of the orbital frequency, and of the
spin parameters. We also study the characteristics of the last stable circular orbit: binding energy,
orbital frequency, and the corresponding dimensionless spin parameter
$\hat{a}_{\rm LSO}\equiv c J_{\rm LSO}/\boldsymbol(G(H_{\rm LSO}/c^2)^2\boldsymbol)$.
We find that the inclusion of NLO spin-orbit terms has a significant ``moderating''
effect on the dynamical characteristics of the circular orbits for large and parallel spins.

\end{abstract}


\maketitle

\section{Introduction}

Coalescing black hole binaries are among the most promising sources
for the currently operating ground-based network
of interferometric detectors of gravitational waves.
It is plausible that the first detections
concern binary systems made of {\it spinning} black holes,
because (as emphasized in \cite{Damour:2001tu}) the spin-orbit
interaction can increase the binding energy of the last stable orbit,
and thereby lead to larger gravitational wave emission.
This makes it urgent to have template waveforms
accurately describing the gravitational wave emission of
spinning binary black holes. These waveforms will be functions
of at least eight intrinsic real parameters: the two masses $m_1$, $m_2$
and the two spin vectors $\mathbf{S}_1$, $\mathbf{S}_2$.
Due to the multi-dimensionality of the parameter space, it seems
impossible for state-of-the-art numerical simulations to densely sample
this parameter space. This gives a clear motivation for developing
{\it analytical} methods for computing the needed, densely spaced,
bank of accurate template waveforms.

Among existing analytical methods for computing the motion
and radiation of binary black hole systems,
the most complete, and the most
promising one, is the Effective One Body (EOB) approach
\cite{Buonanno:1998gg,Buonanno:2000ef,Damour:2000we,Damour:2001tu}.
This method was the first to provide estimates of the complete waveform
(covering inspiral, plunge, merger, and ring-down)
of a coalescing black hole binary, both for non-spinning systems
\cite{Buonanno:2000ef}, and for spinning ones \cite{Buonanno:2005xu}.
Several recent works
\cite{arXiv:0705.2519,arXiv:0706.3732,arXiv:0711.2628,arXiv:0712.3003,DamourNagar08}
have shown that there was an excellent agreement\footnote{For instance,
Ref.\ \cite{arXiv:0712.3003} finds a maximal dephasing
of $\pm0.005$ gravitational wave cycles between EOB and numerical relativity waveforms
describing 12 gravitational wave cycles corresponding to the end of the inspiral,
the plunge, the merger and the beginning of the ringdown
of an equal-mass coalescing binary black hole.}
between the EOB waveforms (for non-spinning systems)
and the results of recent numerical simulations
(see \cite{Pretorius:2007nq} for references and a review
of the recent breakthroughs in numerical relativity).
In addition, the EOB method predicted, before the availability
of reliable numerical relativity (NR) results, a value for the final spin
parameter $\hat{a}_{\rm fin}$ of a coalescing black hole binary
\cite{Buonanno:2000ef,Buonanno:2005xu}
which agrees within $\sim 10\%$ with the results of recent numerical
simulations (see \cite{Pretorius:2007nq} for a review and references).
Recently, it has been shown that the introduction of some
refinements in the EOB approach, led to an EOB/NR agreement for
$\hat{a}_{\rm fin}$ at the $2\%$ level \cite{Damour:2007cb}.

In a previous paper \cite{Damour:2001tu} the EOB method (originally
developed for non-spinning systems) has been generalized to the case
of spinning black holes. It was  shown there that one could map 
the third post-Newtonian (3PN) orbital dynamics, together with the
{\it leading order} (LO) spin-orbit and spin-spin dynamical effects
of a binary system onto an {\it ``effective  test particle'' moving
in a Kerr-type metric}. In the present paper, we extend and refine the
EOB description of spinning binaries by using a recently derived \cite{Damour:2007nc}
{\it Hamiltonian} description of the spin-orbit interaction valid at the
{\it next to leading order} (NLO) in the PN expansion. (The NLO
spin-orbit effects in the harmonic-gauge equations of motion
were first obtained in \cite{Faye:2006gx,Blanchet:2006gy}.)
Let us recall that LO spin-orbit effects are proportional to $G/c^2$,
while NLO ones contain two sorts of contributions:
$\propto G/c^4$ and $\propto G^2/c^4$.
Regarding the spin-spin coupling terms, we shall use here
only the LO results which are made of two different contributions:
the LO $S_1 S_2$ terms \cite{Barker:1970zr}
(which have been recently extended to NLO in \cite{Steinhoff:2007mb}),
and the LO  $S_1^2$ and $S_2^2$ terms. The latter are specific to
Kerr black holes, being related to the quadrupole gravitational
moment of a rotating black hole.\footnote{Note in passing that, if one
wishes to describe the dynamics of, say,
neutron-star binaries with the EOB formalism,
one should add ``correcting'' $S_1^2$ and $S_2^2$ terms.}
It was shown in \cite{Damour:2001tu} that the complete LO spin-spin terms
(the sum of $S_1 S_2$, $S_1^2$, and $S_2^2$ terms)
admitted a remarkable rewriting involving a particular
linear combination $\mathbf{S}_0$, defined below, of the two spin vectors.
This fact, together with the more complicated structure
of spin-orbit terms at the NLO, will lead us below to define a particular,
improved EOB description of spinning binaries.

The present paper consists of two parts:
In the first part (Sections 2 and 3)
we shall develop the formalism needed to finally define
(in Section 4) our improved EOB description of spinning binaries.
In the second part (Section 5), we shall consider one of
the simplest ``applications'' of our EOB Hamiltonian: a
discussion of the energetics of circular, equatorial orbits
for systems with parallel spins. In this section, we shall make contact
with previous related analytical investigations, notably \cite{Blanchet:2006gy},
and prepare the ground for making contact with numerical data.

A few words about our notation: We use the letters $a,b=1,2$ as particle labels.
Then, $m_a$, $\mathbf{x}_a=(x_a^i)$, $\mathbf{p}_a=(p_{ai})$,
and $\mathbf{S}_a=(S_{ai})$ denote, respectively, the mass,
the position vector, the linear momentum vector,
and the spin vector of the $a$th body;
for $a\ne b$ we also define
$\mathbf{r}_{ab}\equiv\mathbf{x}_a-\mathbf{x}_b$,
$r_{ab}\equiv|\mathbf{r}_{ab}|$,
$\mathbf{n}_{ab}\equiv\mathbf{r}_{ab}/r_{ab}$,
$|\cdot|$ stands here for the Euclidean length of a 3-vector.

\section{PN-expanded Hamiltonian}

Our starting point is the PN-expanded (or ``Taylor-expanded'') two-body Hamiltonian $H$
which can be decomposed as the sum of: (i) an orbital part $H_\mathrm{o}$,
(ii) a spin-orbit part $H_\mathrm{so}$ (linear in the spins), and (iii)
a spin-spin term $H_\mathrm{ss}$ (quadratic in the spins),
\begin{align}
\label{fullH}
H(\mathbf{x}_a,\mathbf{p}_a,{\bf S}_a)
&= H_{\mathrm{o}}({\bf x}_a,{\bf p}_a)
+ H_{\mathrm{so}}({\bf x}_a,{\bf p}_a,{\bf S}_a)
\nonumber\\[1ex]&\quad
+ H_{\mathrm{ss}}({\bf x}_a,{\bf p}_a,{\bf S}_a).
\end{align}
The orbital Hamiltonian $H_{\mathrm{o}}$ includes the rest-mass
contribution and is explicitly known (in ADM-like coordinates)
up to the 3PN order \cite{Damour:2000kk,Damour:2001bu}. Its structure is
\begin{align}
\label{horb}
H_{\mathrm{o}}({\bf x}_a,{\bf p}_a) &= \sum_a m_a c^2
+ H_{\text{oN}}({\bf x}_a,{\bf p}_a)
\nonumber\\[1ex]&\quad
+ \frac{1}{c^2}\,H_{\rm o1PN}({\bf x}_a,{\bf p}_a)
+ \frac{1}{c^4}\,H_{\rm o2PN}({\bf x}_a,{\bf p}_a)
\nonumber\\[1ex]&\quad
+ \frac{1}{c^6}\,H_{\rm o3PN}({\bf x}_a,{\bf p}_a)
+ {\cal O}\left(\frac{1}{c^8}\right).
\end{align}
The spin-orbit Hamiltonian $H_\mathrm{so}$ can be written as
\begin{align}
\label{hsot}
H_\mathrm{so}({\bf x}_a,{\bf p}_a,{\bf S}_a)
=\sum_a {\bf\Omega}_{a}({\bf x}_b,{\bf p}_b)\cdot{\bf S}_a,
\end{align}
Here, the quantity ${\bf\Omega}_{a}$ is the sum of a LO contribution
($\propto1/c^2$) and a NLO one ($\propto1/c^4$),
\be
{\bf\Omega}_{a}({\bf x}_b,{\bf p}_b)
= {\bf\Omega}^{\rm LO}_{a}({\bf x}_b,{\bf p}_b)
+ {\bf\Omega}^{\rm NLO}_{a}({\bf x}_b,{\bf p}_b).
\ee
The 3-vectors ${\bf\Omega}^{\rm LO}_{a}$ and ${\bf\Omega}^{\rm NLO}_{a}$
were explicitly computed in Ref.\ \cite{Damour:2007nc}.
They are given, for the particle label $a=1$, by
\begin{widetext}
\begin{subequations}
\label{omegafinal}
\begin{align}
\mathbf{\Omega}^{\rm LO}_{1} &= \frac{G}{c^2r_{12}^2}
\bigg( \frac{3m_2}{2m_1}\ncpi - 2\ncpii \bigg),
\\[2ex]
\mathbf{\Omega}^{\rm NLO}_{1} &= \frac{G^2}{c^4r_{12}^3} \Bigg(
\bigg(-\frac{11}{2}m_2-5\frac{m_2^2}{m_1}\bigg)\ncpi
+ \bigg(6m_1+\frac{15}{2}m_2\bigg)\ncpii \Bigg)
\nonumber\\[1ex]&\quad
+ \frac{G}{c^4r_{12}^2} \Bigg( \bigg(
- \frac{5m_2\pipi}{8m_1^3} - \frac{3\pipii}{4m_1^2}
+ \frac{3\piipii}{4m_1m_2}
- \frac{3\npi\npii}{4m_1^2} - \frac{3\npii^2}{2m_1m_2} \bigg)\ncpi
\nonumber\\[1ex]&\quad
+ \bigg(\frac{\pipii}{m_1m_2}+\frac{3\npi\npii}{m_1m_2}\bigg)\ncpii
+ \bigg( \frac{3\npi}{4m_1^2} - \frac{2\npii}{m_1m_2} \bigg)\cpipii
\Bigg).
\end{align}
\end{subequations}
\end{widetext}
The expressions for $\mathbf{\Omega}^{\rm LO}_{2}$ and
$\mathbf{\Omega}^{\rm NLO}_{2}$ can be obtained from the above formulas by
exchanging the particle labels 1~and~2.

Let us now focus our attention on the dynamics of the {\it relative motion}
of the two-body system in the {\it center-of-mass frame}, which is defined by
the requirement $\mathbf{p}_1+\mathbf{p}_2=\mathbf{0}$. It will be
convenient in the following to work with suitably rescaled variables.
We rescale the phase-space variables
$\mathbf{R}\equiv\mathbf{x}_1-\mathbf{x}_2$ and
$\mathbf{P}\equiv\mathbf{p}_1=-\mathbf{p}_2$
of the relative motion as follows
\be
\label{def1}
{\bf r} \equiv \frac{\mathbf{R}}{GM}
 \equiv \frac{\mathbf{x}_1-\mathbf{x}_2}{GM}, 
\quad
{\bf p} \equiv \frac{\mathbf{P}}{\mu}
\equiv \frac{\mathbf{p}_1}{\mu}
= -\frac{\mathbf{p}_2}{\mu},
\ee
where $M\equiv{m_1+m_2}$ and $\mu\equiv{m_1m_2/M}$.
Note that this change of variables corresponds
to rescaling the action by a factor $1/(GM\mu)$.
It is also convenient to rescale the original time variable $T$ and any
part of the Hamiltonian according to
\be
\label{def1bis}
t \equiv  \frac{T}{GM}, \quad
\hat{H}^{\rm NR} \equiv \frac{H^{\rm NR}}{\mu} ,
\ee
where  $H^{\rm NR} \equiv H - M c^2$ denotes the ``non relativistic''
version of the Hamiltonian, i.e. the Hamiltonian without the rest-mass contribution.
It has the structure
$\hat{H}^{\rm NR}=\frac{1}{2}\pp-\frac{1}{r}+{\cal O}\left(\frac{1}{c^2}\right)$.

It will be convenient in the following to work
with the following two basic combinations of the spin vectors:
\bse
\label{def2}
\begin{align}
\mathbf{S} &\equiv {\bf S}_1 + {\bf S}_2
= m_1 c\,{\bf a}_1 + m_2 c\,{\bf a}_2,
\\[1ex]
\mathbf{S}^* &\equiv \frac{m_2}{m_1}{\bf S}_1
+ \frac{m_1}{m_2}{\bf S}_2
= m_2 c\,{\bf a}_1 + m_1 c\,{\bf a}_2,
\end{align}
\ese
where we have introduced (as is usually done in the
general relativistic literature) the Kerr parameters\footnote{
Note that we use here the usual definition
where the Kerr parameter $a\equiv{S/(Mc)}$
has the dimension of length.
We denote the associated dimensionless rotational parameter
with an overhat: $\hat{a}\equiv a\,c^2/(GM)=c\,S/(GM^2)$.}
of the individual black holes,
$\mathbf{a}_1\equiv\mathbf{S}_1/(m_1c)$ and
$\mathbf{a}_2\equiv\mathbf{S}_2/(m_2c)$.
Note that, in the ``spinning test mass limit'' where, say,
$m_2\to0$ and $S_2\to0$, while keeping $a_2=S_2/(m_2c)$ fixed,
we have a ``background mass'' $M \simeq m_1$, a ``background spin''
${\bf S}_{\rm bckgd}\equiv Mc\,{\bf a}_{\rm bckgd}
\simeq{\bf S}_1=m_1c\,{\bf a}_1$,
a ``test mass'' $\mu \simeq m_2$, and a ``test spin''
$ {\bf S}_{\rm test}={\bf S}_2 =m_2c\,{\bf a}_2 \simeq \mu c\,{\bf a}_{\rm test}$
[with ${\bf a}_{\rm test}\equiv{\bf S}_{\rm test}/(\mu c)$].
Then, in this limit the combination
$\mathbf{S}\simeq\mathbf{S}_1=m_1c\,{\bf a}_1
\simeq Mc\,{\bf a}_{\rm bckgd}={\bf S}_{\rm bckgd}$
measures the background spin, while the other combination,
$\mathbf{S}^*\simeq m_1c\,{\bf a}_2\simeq Mc\,{\bf a}_{\rm test}
=M {\bf S}_{\rm test}/\mu$ measures the (specific) test spin
${\bf a}_{\rm test}={\bf S}_{\rm test}/ (\mu c)$.
The quantities $\mathbf{S}$ and $\mathbf{S}^*$ are the two simplest {\it symmetric}
(under the permutation $1 \leftrightarrow 2$) combinations
of the two spin vectors which have these properties.

In view of the rescaling of the action by a factor $1/(GM\mu)$,
corresponding to the rescaled phase-space variables above, it will
be natural to work with correspondingly rescaled spin variables\footnote{
We recall that (orbital and spin) angular momenta
have the same dimension as the action.}
\be
\label{def3}
\bar{\mathbf{S}}^\mathrm{X}
\equiv \frac{\mathbf{S}^\mathrm{X}}{GM\mu},
\ee
for any label X ($\mathrm{X}=1,2,*, \cdots$).

Making use of the definitions \eqref{def1}--\eqref{def3}
one easily gets from Eqs.\ \eqref{hsot}--\eqref{omegafinal}
the center-of-mass spin-orbit Hamiltonian (divided by $\mu$)
expressed in terms of the rescaled variables:
\begin{align}
\hat{H}_\mathrm{so}(\mathbf{r},\mathbf{p},\bar{\mathbf{S}},\bar{\mathbf{S}}^*)
&= \frac{H_\mathrm{so}(\mathbf{r},\mathbf{p},\bar{\mathbf{S}},\bar{\mathbf{S}}^*)}{\mu}
\nonumber\\[1ex]
&= \frac{1}{c^2}\,\hsoi(\mathbf{r},\mathbf{p},\bar{\mathbf{S}},\bar{\mathbf{S}}^*)
\nonumber\\[1ex]&\qquad
+ \frac{1}{c^4}\,\hsoii(\mathbf{r},\mathbf{p},\bar{\mathbf{S}},\bar{\mathbf{S}}^*)
+ {\cal O}\left(\frac{1}{c^6}\right),
\end{align}
where (here $\mathbf{n}\equiv\mathbf{r}/|r|$)\footnote{
We introduce the following notation for the Euclidean
mixed product of 3-vectors: $(V_1,V_2,V_3)\equiv
\mathbf{V}_1\cdot(\mathbf{V}_2\times\mathbf{V}_3)
=\varepsilon_{ijk}V_1^iV_2^jV_3^k$.}
\begin{widetext}
\bse
\begin{align}
\hsoi(\mathbf{r},\mathbf{p},\bar{\mathbf{S}},\bar{\mathbf{S}}^*)
&= \frac{\nu}{r^2} \Bigg\{ 2\snp + \frac{3}{2}\ssnp \Bigg\},
\\[2ex]
\hsoii(\mathbf{r},\mathbf{p},\bar{\mathbf{S}},\bar{\mathbf{S}}^*)
&= \frac{\nu}{r^3} \Bigg\{ -\left(6+2\nu\right)\snp
- \left(5+2\nu\right)\ssnp \Bigg\}
\nonumber\\[1ex] & \quad
+ \frac{\nu}{r^2} \Bigg\{ \bigglb( \frac{19}{8}\nu\,\pp
+ \frac{3}{2}\nu\,\np^2 \biggrb) \snp
\nonumber\\[1ex] & \qquad\qquad
+ \bigglb( \bigg(-\frac{5}{8}+2\nu\bigg)\pp
+ \frac{3}{4}\nu\,\np^2 \biggrb) \ssnp
\Bigg\},
\end{align}
\ese
\end{widetext}
with $\nu\equiv\mu/M$ ranging from $0$ (test-body limit) to 1/4 (equal-mass case).

Note that the structure of the rescaled spin-orbit Hamiltonian is
\be
\hat{H}_\mathrm{so}(\mathbf{r},\mathbf{p},\bar{\mathbf{S}},\bar{\mathbf{S}}^*)
=\frac{\nu}{ c^2 r^2} \Big( g^\mathrm{ADM}_{S}\snp
+ g^\mathrm{ADM}_{S^*}\ssnp \Big).
\ee
This corresponds to an unrescaled spin-orbit Hamiltonian of the form
\be
{H}_\mathrm{so}= \frac{G}{c^2} \frac{\bf{L}}{R^3} \cdot \Big(
g^\mathrm{ADM}_{S} \mathbf{S} + g^\mathrm{ADM}_{S^*} \mathbf{S}^*
\Big),
\ee
where $R = GMr$ is the unrescaled relative distance (in ADM coordinates),
$\mathbf{L}\equiv\mathbf{R}\times\mathbf{P}=GM\mu\mathbf{r}\times\mathbf{p}$
the relative orbital angular momentum, and where
we have introduced two dimensionless coefficients which might
be  called the ``gyro-gravitomagnetic ratios'', because they
parametrize the coupling between the spin vectors and the ``apparent'' gravitomagnetic
field
$$
\mathbf{v} \times \nabla \frac{G M}{c^2 R}
\propto \frac{ \mathbf{R} \times \mathbf{P}}{R^3}
$$
seen in the rest-frame of a moving particle
(see, e.g., Refs.\ \cite{Damour:1990pi,Damour:1992qi}
for a discussion of the expression of the ``gravitomagnetic field''
in the rest-frame of a moving body).
The explicit expressions of these two gyro-gravitomagnetic ratios are
\begin{widetext}
\bse
\label{gyro}
\begin{align}
g^\mathrm{ADM}_{S}
&= 2 + \frac{1}{c^2} \bigglb(
\frac{19}{8}\nu\,\pp
+ \frac{3}{2}\nu\,\np^2
- \Big(6+2\nu\Big) \frac{1}{r} \biggrb),
\\
g^\mathrm{ADM}_{S^*}
&= \frac{3}{2} + \frac{1}{c^2} \bigglb(
\Big(-\frac{5}{8}+2\nu\Big)\pp
+ \frac{3}{4}\nu\,\np^2
- \Big(5+2\nu\Big) \frac{1}{r} \biggrb).
\end{align}
\ese
\end{widetext}

In the following we shall introduce two related ``effective'' ``gyro-gravitomagnetic ratios'',
that enter the effective EOB Hamiltonian (in effective coordinates).
The label ``ADM'' on the gyro-gravitomagnetic ratios \eqref{gyro}
is a reminder of the fact that the NLO value of these ratios
depend on the precise definition of the radial distance $R$ (which is coordinate dependent).
Let us, however, briefly discuss the origin of the (coordinate-independent)
LO values of these ratios, namely
\be
g^{\mathrm{LO}}_{S} = 2, \quad
g^{\mathrm{LO}}_{S^*} = \frac{3}{2} = 2 - \frac{1}{2}.
\ee
Here the basic ratio 2 which enters both  $g^{\mathrm{LO}}_{S}$
and  $g^{\mathrm{LO}}_{S^*}$ comes from the leading  interaction,
predicted by the Kerr metric,
between the orbital angular momentum of a test particle
and the background spin. See Eq.\ \eqref{mainso} below.
As for the  $-\frac{1}{2}$ ``correction'' in the coupling of the ``test mass'' spin combination
$\bf{S^*}$ it can be seen (e.g.\ from Eq.\ (3.6b) of \cite{Damour:1991rd}) to come from the famous
$\frac{1}{2}$ factor in the Thomas precession (which is a universal, special relativistic
effect, separate from the effects which are specific to the gravitational interaction,
see Eqs.\ (3.2) and (3.3) in \cite{Damour:1991rd}).

To complete this Section, let us recall the remarkable form
[found in Ref.\ \cite{Damour:2001tu}, see Eq.\ (2.54) there]
of the  leading-order spin-spin Hamiltonian $H_\mathrm{ss}$
(including $S_1^2$, $S_2^2$ as well as $S_1 S_2$ terms).
The unrescaled form of the spin-spin Hamiltonian reads
\be
H_\mathrm{ss}(\mathbf{R},{\mathbf{S}}_0)
= \frac{\nu}{2} \frac{G}{c^2} {S}_0^i {S}_0^j
\pa_{ij}\frac{1}{R},
\ee
while its rescaled version reads
\begin{align}
\label{hss}
\hat{H}_\mathrm{ss}(\mathbf{r},\bar{\mathbf{S}}_0)
&\equiv \frac{H_\mathrm{ss}(\mathbf{R},{\mathbf{S}}_0)}{\mu}
\nonumber\\[1ex]
&= \frac{1}{2} \frac{\nu^2}{c^2} \bar{S}_0^i\bar{S}_0^j
\pa_{ij}\frac{1}{r} = \frac{1}{2} \frac{\nu^2}{c^2}
\frac{3(\mathbf{n}\cdot\bar{\mathbf{S}}_0)^2-\bar{\mathbf{S}}_0^2}{r^3}.
\end{align}
The remarkable fact about this result is that it is entirely expressible
in terms of the specific combination of spins
$\mathbf{S}_0 \equiv GM\mu \bar{\mathbf{S}}_0$ defined as:
\be
\label{def2bis}
\mathbf{S}_0 \equiv {\bf S} + {\bf S}^*
= \Big(1 + \frac{m_2}{m_1}\Big){\bf S}_1
+ \Big(1 +\frac{m_1}{m_2}\Big){\bf S}_2.
\ee
We shall come back below to the remarkable properties of the combination $\mathbf{S}_0 $,
which will play a central role in our EOB construction.

\section{Effective Hamiltonian and ``effective gyro-gravitomagnetic'' ratios}

We have obtained in the previous Section the expression of the
full center-of-mass-frame Hamiltonian \eqref{fullH}, in PN-expanded form.
In order to transform this Hamiltonian into a format
which can be resummed in a manner compatible
with previous work on the EOB formalism, we need to perform two
operations on the Hamiltonian (\ref{fullH}). First, we need to
transform the phase-space coordinates $({\bf x}_a,{\bf p}_a,{\bf S}_a)$
by a canonical transformation compatible with the one used in previous
EOB work. Second, we need to compute the {\it effective} Hamiltonian
corresponding to the (canonically transformed) {\it real} Hamiltonian
(\ref{fullH}).

We start by performing the purely orbital canonical transformation
which was found to be needed in Refs.~\cite{Buonanno:1998gg,Damour:2000we}
to go from the ADM coordinates (used in the PN-expanded dynamics)
to the coordinates used in the EOB dynamics. This orbital
canonical transformation is (implicitly) given by
\be
x'^i=x^i+ \frac{\pa G_\mathrm{o} (x,p')}{\pa p'_i},
\quad
p'_i= p_i - \frac{\pa G_\mathrm{o} (x,p')}{\pa x^i}.
\ee
Here the orbital generating function  $G_\mathrm{o} (q,p')$
has been derived to 2PN accuracy in \cite{Buonanno:1998gg},
and to 3PN accuracy in \cite{Damour:2000we}. In the present paper,
as we are only concerned with the additional spin-orbit terms,
treated to 1PN fractional accuracy, it is enough to work with
the 1PN-accurate  generating function  $G_\mathrm{o} (x,p')$. In terms
of  the rescaled variables, the rescaled 1PN-accurate orbital
generating function reads
\begin{align}
\bar{G}_\mathrm{o}({\bf r},{\bf p})
&\equiv \frac{G_\mathrm{o}({\bf r},{\bf p})}{GM\mu}
\nonumber\\[1ex]
&= \frac{1}{c^2} ({\bf r}\cdot{\bf p}) \Bigg( -\frac{1}{2}\nu\,\pp
+ \Big(1+\frac{1}{2}\nu\Big)\frac{1}{r} \Bigg).
\end{align}
This transformation changes the phase-space variables from
$({\bf r},{\bf p},\bar{\mathbf{S}},\bar{\mathbf{S}}^*)$
to $({\bf r}',{\bf p}',\bar{\mathbf{S}},\bar{\mathbf{S}}^*)$.
At the linear order in the transformation (which will be enough for
our purpose), the effect of the transformation on any
of the phase-space variable, say $y$, is
$ y' = y + \{ y,G_\mathrm{o}\}$,
where  $\{\cdot,\cdot\}$ denotes the Poisson bracket.
As $G_\mathrm{o}$ is independent of
time, it leaves the Hamiltonian numerically invariant:
$ H'(y') = H(y)$. This means that it changes the {\it functional}
form of the Hamiltonian according to
$H'(y') = H(y' -\{ y,G_\mathrm{o}\}) = H(y') - \{ H,G_\mathrm{o}\}$.
Note the appearance of the opposite sign in front of the Poisson
bracket, with respect to the effect of the generating function
on the phase-space variables.

As $G_\mathrm{o}$ is of order $1/c^2$, its explicit effect on the two separate terms,
$H^\mathrm{so}_\mathrm{LO}$ and $H^\mathrm{so}_\mathrm{NLO}$, in the
PN expansion of the spin-orbit Hamiltonian is given by:
\bse
\begin{align}
\hsoip({\bf r}',{\bf p}',\bar{\bf S},\bar{\bf S}^*) &=
H^\mathrm{so}_\mathrm{LO}({\bf r}',{\bf p}',\bar{\bf S},\bar{\bf S}^*),
\\[2ex]
\hsoiip({\bf r}',{\bf p}',\bar{\bf S},\bar{\bf S}^*) &=
H^\mathrm{so}_\mathrm{NLO}({\bf r}',{\bf p}',\bar{\bf S},\bar{\bf S}^*)
\nonumber\\[1ex]&\quad
- \{H^\mathrm{so}_\mathrm{LO},\bar{G}_{\mathrm{o}}\}
({\bf r}',{\bf p}',\bar{\bf S},\bar{\bf S}^*).
\end{align}
\ese

It will be convenient in the following to further transform
the phase-space variables by performing a secondary,
purely spin-dependent canonical transformation,
affecting only the NLO spin-orbit terms.
The associated new generating function,
$G_\mathrm{s}({\bf r},{\bf p},\bar{\bf S},\bar{\bf S}^*)$
(assumed to be proportional to the
spins and of order $1/c^4$) will change
the variables $ (y') \equiv ({\bf r}',{\bf p}',\bar{\mathbf{S}},\bar{\mathbf{S}}^*)$
into $(y'') \equiv ({\bf r}'',{\bf p}'',\bar{\mathbf{S}}'',\bar{\mathbf{S}}''^*)$
according to the general rule\footnote{Note that while $G_{\mathrm{o}}$
did not affect the spin variables, the spin-dependent generating
function $G_\mathrm{s}$ will now affect them.} $ y'' = y' + \{ y',G_\mathrm{s}\}$.
For the same reason as above, the (first-order)
effect of $G_\mathrm{s}$ on the functional form of the Hamiltonian
will involve a Poisson bracket with the opposite sign:
$H''(y'') = H(y'') - \{ H,G_\mathrm{s}\}$.

We shall consider a generating function whose unrescaled form reads
\be
{G}_\mathrm{s}({\bf R},{\bf P},{\bf S},{\bf S}^*)
= \frac{G}{\mu\,c^4}\frac{1}{R^3} (\mathbf{R}\cdot\mathbf{P})
(\mathbf{R}\times\mathbf{P})\cdot
\Big( a(\nu)\,\mathbf{S} + b(\nu)\,\mathbf{S}^* \Big),
\ee
while its rescaled  form reads
\begin{align}
\bar{G}_\mathrm{s}({\bf r},{\bf p},\bar{\bf S},\bar{\bf S}^*) &\equiv
\frac{G_\mathrm{s}({\bf R},{\bf P}, {\bf S}, {\bf S}^*)}{GM\mu}
\nonumber\\[1ex]
&= \frac{1}{c^4} \nu \frac{\np}{r} \Bigg( a(\nu) \snp + b(\nu) \ssnp \Bigg).
\end{align}
Here $a(\nu)$ and $b(\nu)$ are two arbitrary, $\nu$-dependent
dimensionless coefficients.\footnote{
The coefficients $a(\nu)$ and $b(\nu)$ can be thought
of as being two ``gauge'' parameters, related to the arbitrariness
in choosing a spin-supplementary condition, and in defining a local
frame to measure the spin vectors.}
Similarly to the result above, the explicit effect
of this new canonical transformation on the two separate terms,
$\hsoip$ and $\hsoiip$, in the PN expansion of the spin-orbit Hamiltonian reads:
\bse
\begin{align}
\hsoipp({\bf r}'',{\bf p}'',\bar{\bf S}'',\bar{\bf S}''^*)
&= \hsoip({\bf r}'',{\bf p}'',\bar{\bf S}'',\bar{\bf S}''^*),
\\[2ex]
\hsoiipp({\bf r}'',{\bf p}'',\bar{\bf S}'',\bar{\bf S}''^*)
&= \hsoiip({\bf r}'',{\bf p}'',\bar{\bf S}'',\bar{\bf S}''^*)
\nonumber\\[1ex]&\quad
- \{H_\mathrm{oN},\bar{G}_{\mathrm{s}}\}
({\bf r}'',{\bf p}'',\bar{\bf S}'',\bar{\bf S}''^*),
\end{align}
\ese
where $H_{\mathrm{oN}}$ is the Newtonian orbital Hamiltonian.
In the following, we shall, for simplicity of notation, omit
the double primes on the new phase-space variables (and on the
corresponding Hamiltonian).

The second operation we need to do is to connect the ``real''
Hamiltonian $H$ to the ``effective'' one $H_\mathrm{eff}$,
which is more closely linked to the description of the
EOB  quasi-geodesic dynamics. The relation between the two
Hamiltonians is quite simple \cite{Buonanno:1998gg,Damour:2000we}:
\be
\label{heff}
\frac{H_\mathrm{eff}}{\mu c^2} \equiv
\frac{H^2 - m_1^2 c^4 - m_2^2 c^4}
{2 m_1 m_2 c^4},
\ee
where we recall that the real Hamiltonian $H$ contains the rest-mass
contribution $Mc^2 =(m_1+m_2)c^2$.
Let us also note that Eq.\ (\ref{heff}) is equivalent to
\be
\frac{H_\mathrm{eff}}{\mu c^2}
= 1 + \frac{H^\mathrm{NR}}{\mu c^2}
+ \frac{1}{2}\nu \frac{(H^\mathrm{NR})^2}{\mu^2 c^4},
\ee
where $H^\mathrm{NR}$ denotes the ``non relativistic'' part of the total
Hamiltonian $H$, i.e., $H^\mathrm{NR}\equiv H-Mc^2$, or more explicitly
\begin{align}
\hnr &= \Big( H_\mathrm{oN} + \frac{H_\mathrm{o1PN}}{c^2}
+ \frac{H_\mathrm{o2PN}}{c^4} + \frac{H_\mathrm{o3PN}}{c^6} \Big)
\nonumber\\[1ex]&\quad
+ \Big( \frac{H^\mathrm{so}_\mathrm{LO}}{c^2}
+ \frac{H^\mathrm{so}_\mathrm{NLO}}{c^4} \Big).
\end{align}
By expanding (in powers of $1/c^2$ and in powers of the spins)
the exact effective Hamiltonian \eqref{heff}, one easily finds that the 
``spin-orbit part'' of the effective Hamiltonian
$H_\mathrm{eff}$ (i.e.\ the part which is linear-in-spin) differs from the 
corresponding part $H_\mathrm{so}$ in the ``real'' Hamiltonian by
a factor $\simeq1+\nu\hat{H}^{\rm NR}/c^2\simeq1+\nu\hat{H}_\mathrm{oN}/c^2$,
so that we get, for the explicit PN expansion of $H_\mathrm{eff}^\mathrm{so}$,
\be
\label{hseff}
\frac{H_\mathrm{eff}^\mathrm{so}}{\mu} = \frac{1}{c^2}\hsoi
+ \frac{1}{c^4}\Big(\hsoii+\nu \hat{H}_\mathrm{oN}\hsoi\Big).
\ee

Combining this result with the effect of the two generating functions
discussed above (and omitting, as we already said, the double primes
on the new phase-space variables
$({\bf r}'',{\bf p}'',\bar{\bf S}'',\bar{\bf S}''^*)$), we get
the transformed spin-orbit part of the effective Hamiltonian in
the form
\be
\label{heffso}
\frac{H_\mathrm{eff}^\mathrm{so}}{\mu}
= \frac{\nu}{c^2 r^2} (\npv) \cdot \Big(  g^\mathrm{eff}_S \bar{\bf S}
+ g^\mathrm{eff}_{S^*} \bar{\bf S}^* \Big),
\ee
which corresponds to the following unrescaled form
(with $\bf{L}\equiv\bf{R}\times\bf{P}$):
\be
\label{heffsounrescaled}
H_\mathrm{eff}^\mathrm{so} = \frac{G}{c^2}
\frac{\bf{L}}{R^3} \cdot \Big(
g^\mathrm{eff}_{S} \mathbf{S}
+ g^\mathrm{eff}_{S^*} \mathbf{S}^* \Big).
\ee
Here the two ``effective gyro-gravitomagnetic'' ratios
$g^\mathrm{eff}_S$ and $g^\mathrm{eff}_{S^*}$ differ from the ``ADM'' ones
introduced above by three effects: (i) a factor
$ \simeq 1 + \nu \hat{H}^{\rm NR}/c^2 \simeq 1 + \nu \hat{H}_\mathrm{oN}/c^2$
due to the transformation from $H$ to $H_\mathrm{eff}$,
(ii) the effect of the orbital  generating function $G_\mathrm{o}$ going from
ADM to EOB coordinates, and (iii) the effect of the spin-dependent 
generating function $G_\mathrm{s}$, which involves the gauge parameters $a(\nu)$
and $b(\nu)$. Their explicit expressions are then found to read
\begin{widetext}
\bse
\label{gyroratios}
\begin{align}
g^\mathrm{eff}_S &\equiv 2 + \frac{1}{c^2}
\bigglb( \Big(\frac{3}{8}\nu+a(\nu)\Big)\pp
- \Big(\frac{9}{2}\nu+3a(\nu)\Big)\np^2
- \frac{1}{r}\Big(\nu+a(\nu)\Big) \biggrb),
\\[2ex]
g^\mathrm{eff}_{S^*} &\equiv \frac{3}{2} +
\frac{1}{c^2} \bigglb( \Big(-\frac{5}{8}+\frac{1}{2}\nu+b(\nu)\Big)\pp
- \Big(\frac{15}{4}\nu+3b(\nu)\Big)\np^2
- \frac{1}{r}\Big(\frac{1}{2}+\frac{5}{4}\nu+b(\nu)\Big) \biggrb).
\end{align}
\ese
\end{widetext}
The choice of the two ``gauge'' parameters
$a(\nu)$ and $b(\nu)$ is arbitrary, and physical results
should not depend on them.\footnote{
Note in particular that the gyro-gravitomagnetic ratios do not
depend on $a(\nu)$ and $b(\nu)$ when considering circular orbits,
i.e.\ when $\pp = 1/r$ and $\np=0$.}
This would be the case if
we were dealing with the exact Hamiltonian. However, as we work
only with an approximation to the exact Hamiltonian, there will
remain some (weak) dependence of our results on the choice of
$a(\nu)$ and $b(\nu)$. We can use this dependence to try to
simplify, and/or to render more accurate, the spin-orbit effects
implied by the above expressions. In particular, we shall focus in this
paper on a special simplifying choice of these gauge parameters:
namely, the values
\be
\label{ab}
a(\nu) = -\frac{3}{8}\nu, \quad
b(\nu) = \frac{5}{8} - \frac{1}{2}\nu,
\ee
which suppress the dependence of the effective gyro-gravitomagnetic
ratios on $\pp$. With this particular choice, the explicit
expressions of these ratios become
\begin{widetext}
\bse
\label{simplegyroratios}
\begin{align}
g^\mathrm{eff}_S &\equiv 2 + \frac{1}{c^2} \bigglb(
- \frac{27}{8}\nu\,\np^2 - \frac{5}{8}\nu\,\frac{1}{r} \biggrb),
\\[2ex]
g^\mathrm{eff}_{S^*} &\equiv \frac{3}{2} +
\frac{1}{c^2} \bigglb(
- \Big(\frac{15}{8}+\frac{9}{4}\nu\Big)\np^2
- \Big(\frac{9}{8}+\frac{3}{4}\nu\Big) \frac{1}{r}\biggrb).
\end{align}
\ese
\end{widetext}

\section{ Spin-dependent Effective-One-Body Hamiltonian}

Up to now we only considered PN-expanded results. In this Section,
we shall generalize the approach of \cite{Damour:2001tu} in
incorporating, in a resummed way, the spin-dependent effects
within the EOB approach. Let us first recall that
the approach of \cite{Damour:2001tu} consists in combining
three different ingredients:
\begin{itemize}
\item a generalized Hamilton-Jacobi
equation involving higher powers of the momenta (as is necessary
at the 3PN accuracy \cite{Damour:2000we});
\item a $\nu$-deformed Kerr-type metric $g^{\alpha\beta}_\mathrm{eff}$,
which depends on the choice of some basic ``effective spin vector''
$S^i_\mathrm{eff}$;
\item the possible consideration of an additional spin-orbit interaction term
$\Delta{H_\mathrm{so}}(\mathbf{r},\mathbf{p},\mathbf{S}_0,\bsigma)$
in the effective Hamiltonian, whose aim is to complete the
spin-dependent interaction incorporated in the definition of
the Hamilton-Jacobi equation based on a certain choice of
``effective spin vector'' $S^i_\mathrm{eff}$.
\end{itemize}

At the LO in spin-orbit and spin-spin interactions,
Ref.\ \cite{Damour:2001tu} showed that one had the choice between
two possibilities:

(i) use as effective spin vector the combination
$\mathbf{S} + \frac{3}{4} \mathbf{S}^*$ which correctly
describes the LO spin-orbit effects, but only approximately
describes the LO spin-spin effects;\footnote{One can then
correct for the missing terms by adding an explicit supplementary
term in the Hamiltonian, quadratic in the spins.} or

(ii) use as effective spin vector the combination
\be
\label{defs0}
\mathbf{S}_0 \equiv {\bf S} + {\bf S}^*
= \Big(1 + \frac{m_2}{m_1}\Big){\bf S}_1
+ \Big(1 +\frac{m_1}{m_2}\Big){\bf S}_2,
\ee
which correctly describes the full LO spin-spin interaction (see
\eqref{hss} above), and complete the description of the LO
spin-orbit effects by adding a term
$\Delta{H_\mathrm{so}}(\mathbf{r},\mathbf{p},\mathbf{S}_0,\bsigma)$
involving a suitably defined spin combination $\bsigma$.
(At LO, Ref.\ \cite{Damour:2001tu} defined
$\bsigma^{\rm LO} = - \frac{1}{4} {\bf S}^*$.)

Intuitively speaking, the second possibility consists in considering
that the ``effective particle'' is endowed not only with a mass $\mu$,
but also with a ``spin'' proportional to $\bsigma$, so that it
interacts with the ``effective background spacetime'' both via
a geodesic-type interaction (described by the
generalized Hamilton-Jacobi equation), and via an additional
spin-dependent interaction proportional to its spin $\propto \bsigma$.

At the present, NLO approximation, where it is crucial to
accurately describe the spin-orbit interaction, as well as,
by consistency, the LO spin-spin ones, we have chosen to
follow the second possibility, which offers more flexibility,
and which looks natural in view of the remarkably simple
LO result \eqref{hss} for the spin-spin interaction (see, however, the
suggestion at the end of the concluding Section 6).

Therefore we shall successively introduce
the ingredients needed to define
\begin{itemize}
\item the Hamilton-Jacobi equation (describing the basic ``geodesic-type''
part of the effective Hamiltonian);
\item the effective, $\nu$-deformed Kerr-type metric $g^{\alpha\beta}_\mathrm{eff}$;
\item the ``effective spin vector'' $S^i_\mathrm{eff}$ entering
the previous Kerr-type metric;
\item the additional spin-orbit interaction
$\Delta{H_\mathrm{so}}(\mathbf{r},\mathbf{p},\mathbf{S}_0,\bsigma)$
involving a new, specific NLO spin combination  $\bsigma$.
\end{itemize}

The modified Hamilton-Jacobi equation \cite{Damour:2000we}
is of the form
\be
\label{HJeq}
g^{\alpha\beta}_\mathrm{eff} P_\alpha P_\beta
+ Q_4(P_i) = -\mu^2 c^2,
\ee
where $Q_4(P_i)$ is a quartic-in-momenta term (which only depends
on the space momentum components $P_i$). For circular orbits
$Q_4(P_i)$ will be zero (see \cite{Damour:2000we,Damour:2001tu}),
so that we will not need its explicit expression in the present paper.

The role of the Hamilton-Jacobi equation above is to allow one to
compute the main part (modulo the additional spin-orbit interaction
added later) of the effective Hamiltonian
$H^{\rm main}_\mathrm{eff}=E_\mathrm{eff}\equiv-P_0c$
by solving \eqref{HJeq} with respect to $P_0$.
The result can be written as
\be
\label{Eeff}
H^{\rm main}_\mathrm{eff} = E_\mathrm{eff}= \beta^i P_i c
+ \alpha c \sqrt{\mu^2 c^2 + \gamma^{ij} P_i P_j + Q_4(P_i)},
\ee
where we have introduced the auxiliary notation
\be
\alpha \equiv (-g^{00}_\mathrm{eff})^{-1/2}, \quad
\beta^i \equiv \frac{g^{0i}_\mathrm{eff}}{g^{00}_\mathrm{eff}},
\quad \gamma^{ij} \equiv g^{ij}_\mathrm{eff}
- \frac{g^{0i}_\mathrm{eff}\,g^{0j}_\mathrm{eff}}{g^{00}_\mathrm{eff}}.
\ee

The next crucial ingredient consists in defining the  (spin-dependent)
effective metric entering the Hamilton-Jacobi equation,
and thereby the effective Hamiltonian \eqref{Eeff}.
We shall follow here Ref.\ \cite{Damour:2001tu}
in employing an effective co-metric of the form
(here $P_t\equiv cP_0$)
\begin{align}
\label{cometric1}
g^{\alpha\beta}_\mathrm{eff} P_{\alpha} P_{\beta}
&= \frac{1}{R^2+a^2\cos^2\theta}
\bigglb( \Delta_R(R)\,P_R^2 + P_\theta^2
\nonumber\\[1ex]&\quad
+ \frac{1}{\sin^2\theta}\Big(P_\phi+a\sin^2\theta\frac{P_t}{c}\Big)^2
\nonumber\\[1ex]&\quad -\frac{1}{\Delta_t(R)}
\Big((R^2+a^2)\frac{P_t}{c}+ a\,P_\phi\Big)^2 \biggrb),
\end{align}
where the functions $\Delta_t$ and $\Delta_R$ are defined as
\bse
\begin{align}
\label{deltat}
\Delta_t(R) &\equiv R^2 P^n_m\Big[A(R)+\frac{a^2}{R^2}\Big],
\\[1ex]
\label{deltar}
\Delta_R(R) &\equiv \Delta_t(R) D^{-1}(R),
\end{align}
\ese
and where  the Kerr-like parameter $a$
is defined as $a\equiv S_\mathrm{eff}/(Mc)$,
where $S_\mathrm{eff}$ denotes the modulus of the
``effective spin vector'' $S^i_\mathrm{eff}$
entering the definition of the
Kerr-like metric above. We shall come back below to
the choice of this vector $S^i_\mathrm{eff}$ (which is
one of the ingredients in the definition of a
spin-dependent EOB formalism).
In Eq.\ \eqref{deltat} $P^n_m$ denotes the operation
of taking the $(n,m)$-Pad\'e approximant,\footnote{
Let us recall that the $(n,m)$-Pad\'e approximant
of a function $c_0+c_1u+c_2u^2+\cdots+c_{n+m}u^{n+m}$
is equal to $N_n(u)/D_m(u)$,
where $N_n(u)$ and $D_m(u)$ are polynomials
in $u$ of degrees $n$ and $m$, respectively.}
and the PN expansions of the metric coefficients $A$ and $D^{-1}$ equal
(here $\hat{u}\equiv GM/(Rc^2)$)
\bse
\begin{align}
\label{Aeff}
A(\hat{u}) &= 1 - 2\hat{u} + 2\nu\hat{u}^3
+ \Big(\frac{94}{3}-\frac{41}{32}\pi^2\Big)\nu\hat{u}^4,
\\[1ex]
D^{-1}(\hat{u}) &= 1 + 6\nu\hat{u}^2 + 2(26-3\nu)\nu\hat{u}^3.
\end{align}
\ese

For pedagogical clarity,
we have given above the expression of the effective EOB metric
in a Boyer-Lindquist-type coordinate system
aligned with the instantaneous direction of the (time-dependent)
effective spin vector $S^i_\mathrm{eff}$. This expression
will suffice in the present paper where we will only consider
situations where the spin vectors are aligned with the orbital
angular momentum, so that they are fixed in space. As emphasized
in \cite{Damour:2001tu}, when applying the EOB formalism to
more general situations (non aligned spins) one must
rewrite the effective co-metric components in a ``fixed''
Cartesian-like coordinate system. This is done by introducing
\begin{align}
n^i &\equiv x^i/R, \quad
s^i \equiv \frac{S^i_\mathrm{eff}}{S_\mathrm{eff}}, \quad
\cos\theta \equiv n^i s^i,
\nonumber\\[1ex]
\rho &\equiv \sqrt{R^2 + a^2 \cos^2\theta},
\end{align}
and rewriting the co-metric components as
\bse
\label{cometric2}
\begin{align}
g^{00}_\mathrm{eff}
&= -\frac{(R^2+a^2)^2-a^2\Delta_t(R)\sin^2\theta}
{\rho^2\,\Delta_t(R)},
\\[1ex]
g^{0i}_\mathrm{eff}
& = -\frac{a(R^2+a^2-\Delta_t(R))}{\rho^2\,\Delta_t(R)}
(\mathbf{s}\times\mathbf{R})^i,
\\[1ex]
g^{ij}_\mathrm{eff} &= \frac{1}{\rho^2}\Big(\Delta_R(R)n^in^j
+ R^2(\delta^{ij}-n^in^j)\Big)
\nonumber\\[1ex]&\quad
- \frac{a^2}{\rho^2\,\Delta_t(R)}
(\mathbf{s}\times\mathbf{R})^i(\mathbf{s}\times\mathbf{R})^j.
\end{align}
\ese
Making use of Eqs.\ \eqref{cometric2} one computes
\bse
\begin{align}
\alpha &= \rho
\sqrt{\frac{\Delta_t(R)}{(R^2+a^2)^2-a^2 \Delta_t(R) \sin^2\theta}},
\\[1ex]
\beta^i &=
\frac{a(R^2+a^2-\Delta_t(R))}{(R^2+a^2)^2-a^2\Delta_t(R)\sin^2\theta}
(\mathbf{s}\times\mathbf{R})^i,
\\[1ex]
\gamma^{ij} &= g^{ij}_\mathrm{eff} + \frac{\beta^i\beta^j}{\alpha^2}.
\end{align}
\ese
Replacing the latter expressions in the general form
of the effective energy \eqref{Eeff} yields the most general form
of the main part of the effective Hamiltonian
$H^\mathrm{main}_\mathrm{eff}(\mathbf{x},\mathbf{P},\mathbf{S}_a)$.

The definition of
$H^\mathrm{main}_\mathrm{eff}(\mathbf{x},\mathbf{P},\mathbf{S}_a)$
crucially depends on the choice of effective Kerr-type spin vector.
In order to automatically incorporate, in a correct manner, the LO
spin-spin terms, we shall use here
\be
\label{defseff}
Mc\,\mathbf{a} \equiv
\mathbf{S}_{\rm eff} \equiv \mathbf{S}_0 = {\bf S} + {\bf S}^*
= \Big(1 + \frac{m_2}{m_1}\Big){\bf S}_1
+ \Big(1 +\frac{m_1}{m_2}\Big){\bf S}_2.
\ee
Note that, besides its usefulness in treating spin-spin effects,
this definition has several nice features. For example, if
we introduce the Kerr parameters
of the individual black holes,
$ \mathbf{a}_1\equiv\mathbf{S}_1/(Mc)$,
$ \mathbf{a}_2\equiv\mathbf{S}_2/(Mc)$,
the Kerr parameter $\mathbf{a}_0\equiv\mathbf{S}_0/(Mc)$
(where we naturally take $m_0=M=m_1+m_2$)
associated to the spin combination \eqref{defs0} is simply
\be
\mathbf{a}_0 =\mathbf{a}_1 + \mathbf{a}_2.
\ee
Let us also note that the corresponding {\it dimensionless}
spin parameters (with, again, $m_0 = M = m_1 + m_2$)
\be
\hat{\bf a}_i \equiv \frac{c\,{\bf S}_i}{G m_i^2},
\quad i=0,1,2,
\ee
satisfy
\be
\hat{\bf a}_0 = X_1 \hat{\bf a}_1 + X_2 \hat{\bf a}_2,
\ee
where $X_1 \equiv m_1/M$ and $X_2 \equiv m_2/M$ are the
two dimensionless mass ratios (with $X_1 + X_2 = 1$ and
$X_1 X_2 = \nu$). This last result shows that, in
$\hat{\bf a}$-space, the ``point'' $\hat{\bf a}_0$ is
on the straight-line segment joining the two ``points''
$\hat{\bf a}_1$ and $\hat{\bf a}_2$. The individual
Kerr bounds tell us that each point $\hat{\bf a}_1$ and $\hat{\bf a}_2$
is contained within the unit Euclidean sphere. By convexity of the
unit ball, we conclude that the ``effective'' dimensionles
spin parameter $\hat{\bf a}_0$ will also automatically satisfy
the Kerr bound $|\hat{\bf a}_0| \leq 1$. This is a nice
consistency feature of the definition of the associated
Kerr-type metric.

It remains to define the additional ``test-spin''
vector $\bsigma$, and the associated
additional effective spin-orbit interaction
term. Following the logic of \cite{Damour:2001tu}
(and generalizing the LO results given in Eqs.\ (2.56)--(2.58) there),
these quantities are defined by
\begin{align}
\label{sigma}
\bsigma &\equiv \frac{1}{2} g^\mathrm{eff}_S \mathbf{S}
+ \frac{1}{2}g^\mathrm{eff}_{S^*}\mathbf{S}^* - \mathbf{S}_{\rm eff}
\nonumber\\[1ex]
&= \frac{1}{2}\big(g^\mathrm{eff}_S-2\big)\mathbf{S}
+ \frac{1}{2}\big(g^\mathrm{eff}_{S^*}-2\big)\mathbf{S}^*,
\end{align}
and
\begin{widetext}
\be
\Delta{H_\mathrm{so}}(\mathbf{x},\mathbf{P},\mathbf{S}_0,\bsigma)
\equiv \frac{R^2+a_0^2-\Delta_t(R)}
{(R^2+a_0^2)^2-a_0^2\Delta_t(R)\sin^2\theta_0}
\frac{(P,\sigma,R)}{M},
\ee
where $\mathbf{a}_0\equiv\mathbf{S}_0/(Mc)$
and $\cos\theta_0\equiv{n^iS^i_0}/|\mathbf{S}_0|$.
The justification for these definitions is that
the ``main'' Hamilton-Jacobi part of the effective Hamiltonian
contains, as spin-orbit (i.e.\ linear-in-spin) part,
the following term
\begin{align}
\label{mainso}
H^\text{main\,eff}_\mathrm{so}
&= c P_i \big(\beta^i \big)_\text{linear-in-spin}
\nonumber\\[1ex]
&= c P_i \bigg(\frac{R^2+a_0^2-\Delta_t(R)}
{(R^2+a_0^2)^2-a_0^2\Delta_t(R)\sin^2\theta_0}
(\mathbf{a}_0\times\mathbf{R})^i\bigg)_\text{linear-in-spin}
\nonumber\\[1ex]
&= \frac{2GM}{c R^3} P_i (\mathbf{a}_0\times\mathbf{R})^i
+ (\text{NNLO corrections})
\nonumber\\[1ex]
&= 2  \frac{G}{c^2} \frac{\bf{L}}{R^3} \cdot \mathbf{S}_0
+ (\text{NNLO corrections}),
\end{align}
\end{widetext}
where the factor $2GM$ comes from the second term in the PN expansion of
$\Delta_t(R) = R^2 - 2 GM R/c^2 + 2 \nu (GM)^3/(R\,c^6)
+ \text{(quadratic-in-spin terms)}$.
Note that the absence of $c^{-4}$ correction in the effective metric function
$A(R)$ means that the leading term $\propto2GM$ in the spin-orbit
part of $H^{\rm main}$ is valid {\it both} to LO and to NLO,
i.e., up to ``next to next to leading order'' (NNLO).

When comparing this result to the NLO result \eqref{heffsounrescaled},
we see that the ``main'' part of the effective Hamiltonian
contains a spin-orbit piece which is equivalent
to having effective gyro-gravitomagnetic ratios equal to
$g^\text{main\,eff}_S = 2$ and $g^\text{main\,eff}_{S^*} = 2$,
instead of the correct values derived above. One then easily checks that
the definition above of $\bsigma$ and of the associated
supplementary spin-orbit interaction
$\Delta{H_\mathrm{so}}(\mathbf{x},\mathbf{P},\mathbf{S}_0,\bsigma)$
has the effect of including the full result for the
NLO spin-orbit interaction. It is also to be noted that the additional spin-orbit 
interaction $\Delta{H_\mathrm{so}}$ goes to zero  proportionally to $\nu$ in
the test mass limit $m_2 \to 0$ because, on the one hand,
$g^\mathrm{eff}_S-2$ is proportional to $\nu$ (if $a(\nu)$ is),
and, on the other hand, though $g^\mathrm{eff}_{S^*}-2$
{\it does not} tend to zero with $\nu$,
the second spin combination $\mathbf{S}^*$
{\it does} tend to zero proportionally to $\nu$
[see Eqs.\ \eqref{aa*} below].

Summarizing: we propose to define
a total effective spin-dependent Hamiltonian of the form
\be
\label{hteff}
H_\mathrm{eff}(\mathbf{x},\mathbf{P},\mathbf{S}_1,\mathbf{S}_2)
\equiv {H}^{\rm main}_{\mathrm{eff}}(\mathbf{x},\mathbf{P},\mathbf{S}_0)
+ \Delta{H_\mathrm{so}}(\mathbf{x},\mathbf{P},\mathbf{S}_0,\bsigma),
\ee
where ${H}^{\rm main}_{\mathrm{eff}}(\mathbf{x},\mathbf{P},\mathbf{S}_0)$
is given by the right-hand side of Eq.\ \eqref{Eeff}
computed for the effective spin variable equal to
$\mathbf{S}_0$ [defined in Eq.\ \eqref{defs0}] and where
$\Delta{H_\mathrm{so}}(\mathbf{x},\mathbf{P},\mathbf{S}_0,\bsigma)$
is the additional spin-orbit interaction term defined above
[with $\mathbf{a}_0\equiv\mathbf{S}_0/(Mc)$].

Finally, the {\em real EOB-improved} Hamiltonian
(by contrast to the ``effective'' one)
is defined by solving Eq.\ \eqref{heff} with respect to
$H_\mathrm{real}=H^\mathrm{NR}+Mc^2$:
\be
\label{hreal}
H_\mathrm{real} = Mc^2
\sqrt{1+2\nu\Big(\frac{H_\mathrm{eff}}{\mu c^2}-1\Big)},
\ee
where $H_\mathrm{eff}$ is given in Eq.\ \eqref{hteff}.

\section{Dynamics of circular orbits}

In this Section we shall apply the construction of the
NLO spin-dependent EOB Hamiltonian to the study of the dynamics
of circular orbits of binary black hole systems.

Besides the dimensionless spin parameters $\hat{\bf a}_1$
and $\hat{\bf a}_2$ already introduced above,
it is convenient to introduce the dimensionless spin variables
corresponding to the basic spin combinations
${\bf S}$ and ${\bf S}^*$, namely
\be
\hat{\bf a} \equiv \frac{c\,{\bf S}}{GM^2},
\quad
\hat{\bf a}^* \equiv \frac{c\,{\bf S}^*}{GM^2}.
\ee
Let us note in passing the various links between the
dimensionless spin parameters that one can define
[including $\hat{\bf a}_0\equiv{c\,{\bf S}_0}/(GM^2)$
already introduced above],
\bse
\label{aa*}
\begin{gather}
\hat{\bf a} = X_1^2 \hat{\bf a}_1 +  X_2^2 \hat{\bf a}_2,
\quad
\hat{\bf a}^* = \nu  \hat{\bf a}_1 + \nu \hat{\bf a}_2,
\\[1ex]
\hat{\bf a}_0 = \hat{\bf a} + \hat{\bf a}^*= X_1\hat{\bf a}_1
+X_2 \hat{\bf a}_2.
\end{gather}
\ese
Here as above we use the mass ratios
$X_1 \equiv m_1/M$, $X_2 \equiv m_2/M$
such that $X_1 + X_2 =1$ and $X_1 X_2=\nu$.
Let us note that for equal-mass binaries
($m_1=m_2$, $X_1=X_2=\frac{1}{2}$),
with arbitrary (possibly unequal) spins, one has
$\hat{\bf a}=\hat{\bf a}^*
=\frac{1}{4} (\hat{\bf a}_1+\hat{\bf a}_2)
=\frac{1}{2}\hat{\bf a}_0$.
Note also that, in the test-mass limit, say $m_1 \gg m_2$
so that  $X_1 \to 1$ and $ X_2 \to 0$, one has
\be
\hat{\bf a} = \hat{\bf a}_0 = \hat{\bf a}_1,
\quad \hat{\bf a}^* = 0.
\ee

In the general case where the spin vectors are not aligned
with the (rescaled) orbital angular momentum vector\footnote{
In the following, we switch again to the use of scaled variables:
$\mathbf{r}\equiv{\mathbf{R}/(GM)}$,
$\bl\equiv{\mathbf{L}/(GM\mu)}$, and
$\mathbf{p}\equiv\mathbf{P}/\mu$.} $\bl$,
\be
\bl = r\,\mathbf{n}\times\mathbf{p},
\ee
there exist no circular orbits. However, there
exist (at least to a good approximation) some ``spherical
orbits'', i.e.\ orbits that keep a constant value of the
modulus of the radius vector $\bf{r}$, though they do not
stay within one fixed plane. As discussed in \cite{Damour:2001tu}
one can analytically study these spherical orbits within the
EOB approach, and discuss, in particular, the characteristics
of the {\em last stable spherical orbit}.

For simplicity, we shall restrict ourselves here to the situation
where both individual spins are parallel (or antiparallel)
to the orbital angular momentum vector $\bl$. In that case,
we can consistently set everywhere the radial momentum to zero,
$p_r =\mathbf{n}\cdot\mathbf{p} = 0$, and express the (real)
EOB Hamiltonian as a function of $r$, $\ell = p_{\varphi}$
(using  $\mathbf{p}^2 = \ell^2/r^2$, where $\ell\equiv|\bl|$),
and of the two scalars $\hat{a}, \hat{a}^*$
measuring the projections of our basic spin combinations on
the direction of the orbital angular momentum $\bl$. They are
such that
\begin{align}
\hat{\mathbf{a}}\cdot\bl = \hat{a}\,\ell, \quad
\hat{\mathbf{a}}^*\cdot\bl = \hat{a}^*\,\ell.
\end{align}
The scalars $\hat{a}$ and $\hat{a}^*$ can be either positive or negative,
depending on whether, say, $\hat{\mathbf{a}}$ is parallel or antiparallel to $\bl$.

The sequence of circular (equatorial) orbits is then determined by the
constraint
\be
\frac{\pa H_\mathrm{real}(r,\ell,\hat{a},\hat{a}^*)}{\pa r} = 0.
\ee
Then, the angular velocity along each circular orbit is
given by
\be
\Omega \equiv \frac{1}{GM\mu}
\frac{\pa H_\mathrm{real}(r,\ell,\hat{a},\hat{a}^*)}{\pa\ell}.
\ee
As mentioned above, we have chosen the special
values  $a(\nu) = -\frac{3}{8}\nu$,
$b(\nu) = \frac{5}{8} - \frac{1}{2}\nu$
of the two gauge parameters, to simplify the expression of the Hamiltonian.

In Figs.\ 1--4 we explore several aspects of the dynamics of circular orbits,
using as basic diagnostic the relation between the energy and the angular velocity
along the sequence of circular orbits (``binding energy curve'').
More precisely, we plot the dimensionless ``non relativistic'' energy
\be
e \equiv \frac{H_\mathrm{real}}{Mc^2} - 1,
\ee
as a function of the dimensionless angular velocity:
\be
\hat{\Omega} \equiv \frac{GM}{c^3} \Omega.
\ee

\begin{figure}
\scalebox{0.7}{\includegraphics{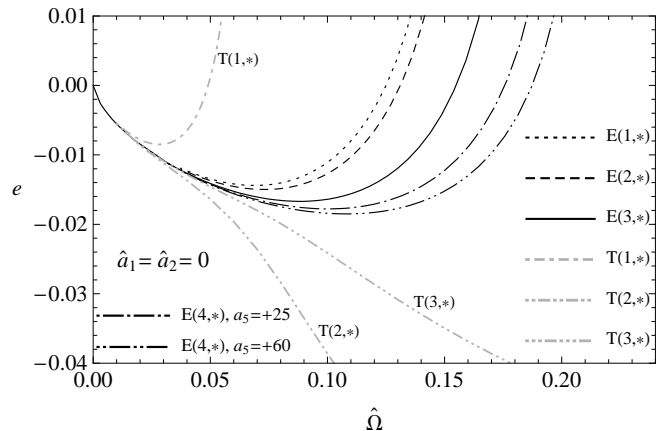}}
\caption{
Binding energy curves for circular orbits of symmetric
{\it non-spinning} binaries ($m_1=m_2$
and $\hat{\mathbf{a}}_1=\hat{\mathbf{a}}_2=\mathbf{0}$):
dimensionless ``non relativistic'' energy $e$
versus dimensionless angular frequency $\hat{\Omega}$.
The notation E$(n,*)$ means computation of the energy
using the EOB-improved real Hamiltonian \eqref{hreal}
with the $n$PN-accurate metric function $\Delta_t(R)$;
the function $\Delta_t(R)$ was computed
by means of Eq.\ \eqref{deltat}
using the $(1,n)$ Pad\'e approximant  at the $n$PN order.
Here $n=1,2,3,4$, where $n=4$ refers to the ``4PN'' case where
a term $+a_5\nu\hat{u}^5$  is added to the function $A(\hat{u})$.
For the curves labelled by T$(n,*)$ the computation was done with the direct
PN-expanded (ADM-coordinates) orbital Hamiltonian \eqref{horb}
with the terms up to the $n$PN order included.}
\end{figure}

For simplicity, we shall restrict most of our studies to {\it symmetric}
binary systems, i.e.\ systems with $m_1=m_2$ and $a_1=a_2$. For such systems
the dimensionless effective spin parameter is $\hat{a}_0=\hat{a}_1=\hat{a}_2$.
The information contained in these figures deals with the following aspects of
the description of the dynamics:

\begin{itemize}

\item As a warm up, and a reminder,
Fig.\ 1 considers the case of {\it non-spinning binaries} (i.e.\ $\hat{a}_0=0$).
This figure contrasts  the behaviour of the successive PN versions of the EOB dynamics, 
with that of the successive PN versions of the non-resummed,
``Taylor-expanded'' Hamiltonian. The numbers 1,2,3 refer to 1PN, 2PN, and 3PN,
while the letter ``E'' refers to ``EOB''
and the letter ``T'' refers to ``Taylor''. For instance, E$(3,*)$ refers to the
$e(\hat{\Omega})$ binding energy curve computed with the 3PN-accurate EOB
Hamiltonian. [The star in E$(3,*)$ replaces the label we shall use below to
distinguish LO versus NLO treatment of spin-orbit effects. In the present
non-spinning case we are insensitive to this distinction.]
To be precise, the notation E$(n,*)$ refers to a computation
of the circular orbits using the $\hat{a}_0\to0$ limit\footnote{
Note that the $\hat{a}_0 \to 0$ limit of the
Pad\'e  resummation of some $\hat{a}_0$-dependent metric coefficient
is not necessarily the same as the Pad\'e approximant
one might normally consider in the non-spinning case.}
of the EOB-improved real Hamiltonian \eqref{hreal} with the $n$PN-accurate
metric function $\Delta_t(R)$; where $\Delta_t(R)$ was computed
by means of Eq.\ \eqref{deltar}
using the following Pad\'e approximants:
(1,1) at the 1PN order,
(1,2) at the 2PN order, and
(1,3) at the 3PN order.
As for the Taylor-based approximants to the binding energy curve, T$(n,*)$,
they were computed by using as basic Hamiltonian (to
define the dynamics) the $n$PN-accurate Taylor-expanded
Hamiltonian, in ADM coordinates, \eqref{horb},
without doing any later PN re-expansion.\footnote{
As is well-known there are always many non-equivalent ways
of defining any ``$n$PN'' result, depending of where, and how,
in the calculation one is replacing a function by a PN-expanded polynomial.
For instance, one could PN re-expand the function giving
the energy $e$ in terms of the orbital frequency $\hat{\Omega}$,
or the function giving $e$ in terms of the orbital angular momentum $L$
(see Ref.\ \cite{Damour:2000we} for the computation of several such
functions in the non-spinning case). However,
we are ultimately interested (for gravitational-wave purposes)
in defining a complete dynamics for coalescing spinning binaries.
Therefore, we focus here on the results
predicted by Hamiltonian functions $H(x,p,\cdots)$.}

It is interesting to note
that the successive PN-approximated EOB binding energy curves
are stacked in a {\it monotonically  decreasing} fashion,
when increasing the PN accuracy,
and all admit a minimum at some value of the orbital frequency.
This minimum corresponds to the last stable circular orbit (see below).
The monotonic stacking of the EOB energy curves therefore implies
that a higher PN accuracy predicts circular orbits
which are more bound, and can reach higher orbital frequencies.
Let us note in this respect that
recent comparisons between EOB and numerical relativity data
have found the need to add
a {\it positive} 4PN additional term $+a_5\nu\hat{u}^5$
in the basic EOB radial potential $A(\hat{u})$ of Eq.\ \eqref{Aeff} above,
with $a_5$ somewhere between $+10$ and $+80$
\cite{arXiv:0706.3732,arXiv:0711.2628,arXiv:0712.3003,DamourNagar08}.
Though we do not know yet
what is the ``real'' value of the 4PN coefficient $a_5$
we have included in Fig.\ 1 two illustrative\footnote{
These two values of the 4PN parameter $a_5$
were found in Refs.\ \cite{arXiv:0706.3732,arXiv:0712.3003}
to be representative of the values of $a_5$
that improve the agreement between EOB waveforms and numerical relativity ones.}
values of this ``4PN'' orbital parameter,
namely $a_5= +25$ and  $a_5= +60$.
Note that the effect of such positive values of $a_5$
is to push the last few circular orbits towards more bound,
higher orbital frequency orbits. This effect will compound itself with
the effects of spin explored below, and should be kept in mind when
looking at our other plots.

By contrast with the ``tame'' and monotonic behaviour of successive
EOB approximants, we see on Fig.\ 1 that the successive Taylor-Hamiltonian
approximants T$(n,*)$ have a more erratic behaviour. Note in particular, that
the 3PN-accurate energy curve does not admit any minimum as the
orbital frequency increases (in other words, there is no ``last'' stable
circular orbit). In view of this bad behaviour of the 3PN-accurate
orbital Taylor-Hamiltonian, we shall not consider anymore
in the following figures the predictions coming
from such Taylor Hamiltonians.\footnote{
Indeed, in the physically most important case of
parallel (rather than anti-parallel) spins, the spin-orbit coupling
will be repulsive (like the effect of a positive $a_5$), and will
tend to reinforce the ``bad'' behaviour of the 3PN orbital Taylor
Hamiltonian (i.e.\ the absence of any last stable orbit).}

\begin{figure}
\scalebox{0.7}{\includegraphics{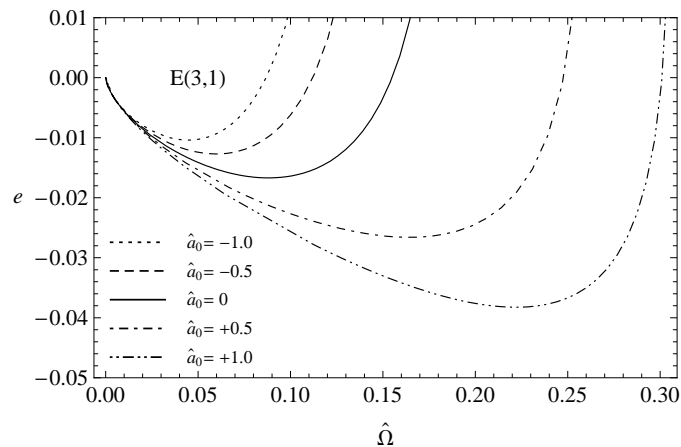}}
\caption{Binding energy curves for circular orbits
of symmetric {\it parallely spinning} binaries ($m_1=m_2$
and $\hat{\mathbf{a}}_1=\hat{\mathbf{a}}_2 \propto\bf{ r}\times \bf{p}$):
dimensionless energy $e$ versus dimensionless angular frequency $\hat{\Omega}$
along circular orbits for various values of the dimensionless effective spin
parameter $\hat{a}_0\equiv{c\,{\bf S}_0}/(GM^2)= \hat{a}_1=\hat{a}_2$
within the effective-one-body approach.
The label E$(3,1)$ means that we use the EOB Hamiltonian with 3PN-accurate
orbital effects and NLO spin-orbit coupling, i.e.\ Eq.\ \eqref{sigma} was used
with the NLO gyro-gravitomagnetic ratios
$g^\mathrm{eff}_S$ and $g^\mathrm{eff}_{S^*}$, Eqs.\ \eqref{gyroratios}.}
\end{figure}

\item In Fig.\ 2 we study the effect of changing the amount of spin on
the black holes of our binary system. We use here our new, NLO spin-orbit
EOB Hamiltonian, as indicated by the notation E$(3,1)$, where the first label, 3,
refers to the 3PN accuracy, and the second label, 1, to the 1PN fractional accuracy of
the spin-orbit terms (i.e., the NLO accuracy).
Note that the EOB binding energy curves are stacked
in a monotonically decreasing way as the dimensionless effective spin $\hat{a}_0$
increases from  $\hat{a}_0= -1 $ (maximal spins antiparallel to the orbital
angular momentum) to $\hat{a}_0= +1 $ (maximal spins parallel to the orbital
angular momentum). Note also that this curve confirms the finding of \cite{Damour:2001tu}
that parallel spins lead to the possibility of closer and more bound circular orbits.

\item Fig.\ 3 contrasts the effect of using the NLO spin-orbit interaction
instead of the LO one in the EOB Hamiltonian.
 We use the full 3PN accuracy,
and include the LO spin-spin interaction.
E$(3,0)$ denotes a result obtained with the 3PN-accurate EOB
Hamiltonian using the LO (or 0PN-accurate) spin-orbit terms, while
E$(3,1)$ uses the  3PN-accurate EOB Hamiltonian with NLO (1PN-accurate)
spin-orbit terms. Each panel in the Figure corresponds to a specific value
of the dimensionless effective spin $\hat{a}_0$. To guide the eye we use
in all our figures a solid line to denote our ``best'' description, i.e.
the 3PN-NLO EOB E$(3,1)$. Note that the addition of the NLO
effects in the spin-orbit interaction has the clear effect of {\it moderating}
the influence of the spins (especially for positive spins).
While the binding energy curves using the LO spin-orbit effects
tend to abruptly dive down towards very negative
energies when the spins are large and positive,\footnote{As
discussed in Section 3C of Ref.\ \cite{Damour:2001tu}, this is due to
the then {\it repulsive} character of the spin-orbit (and spin-spin)
interaction.}
the corresponding
NLO curves have a much more moderate behaviour.

\end{itemize}

\begin{figure*}
\begin{center}
\begin{tabular}{cc}
\scalebox{0.65}{\includegraphics{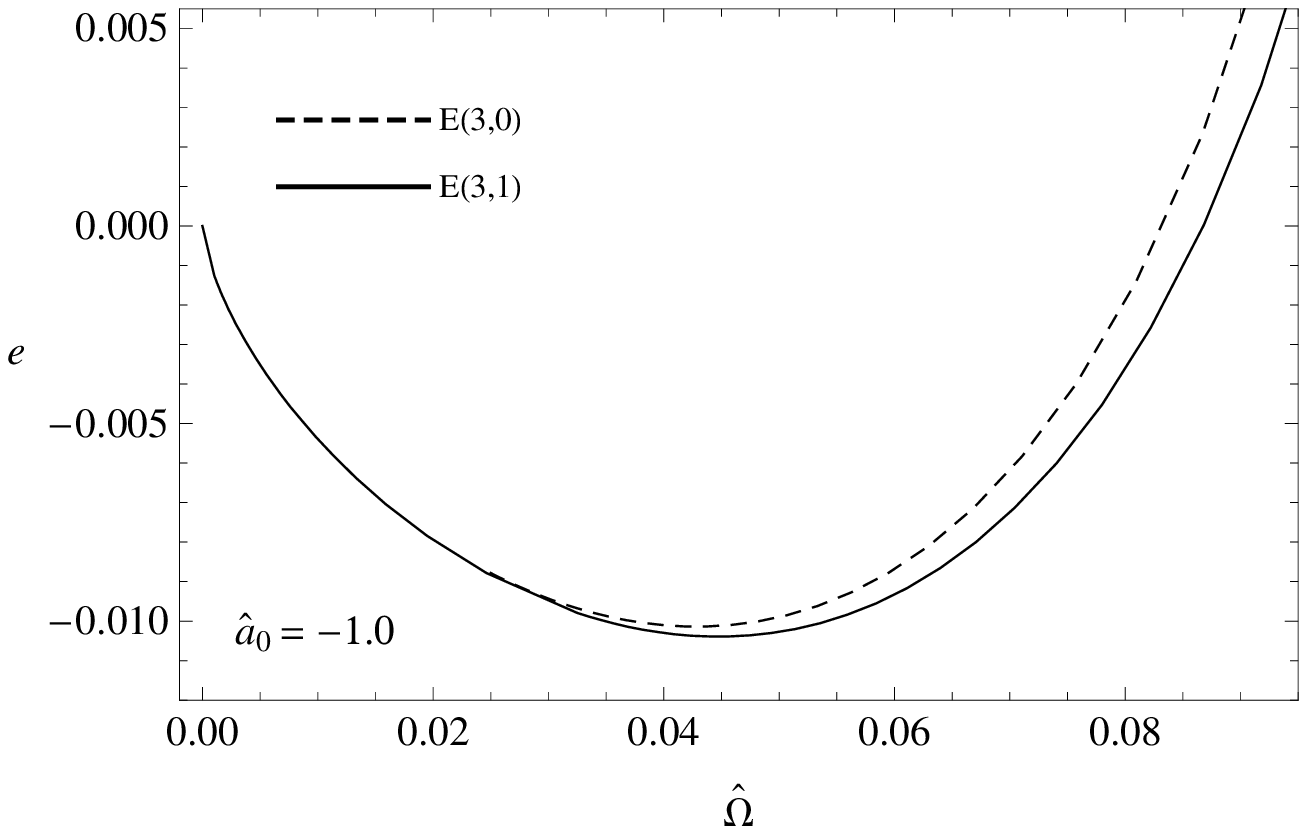}}&
\scalebox{0.65}{\includegraphics{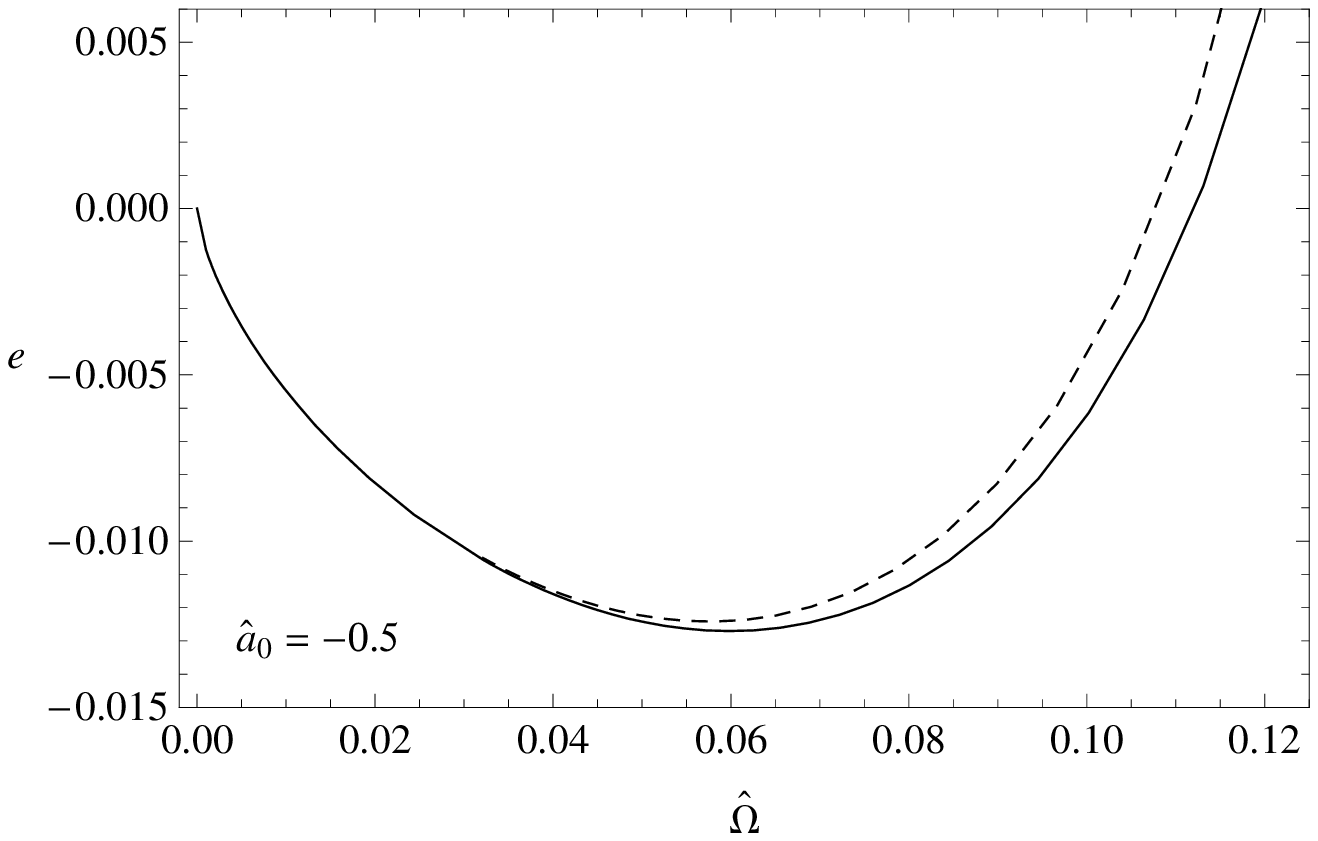}}\\[2ex]
\scalebox{0.65}{\includegraphics{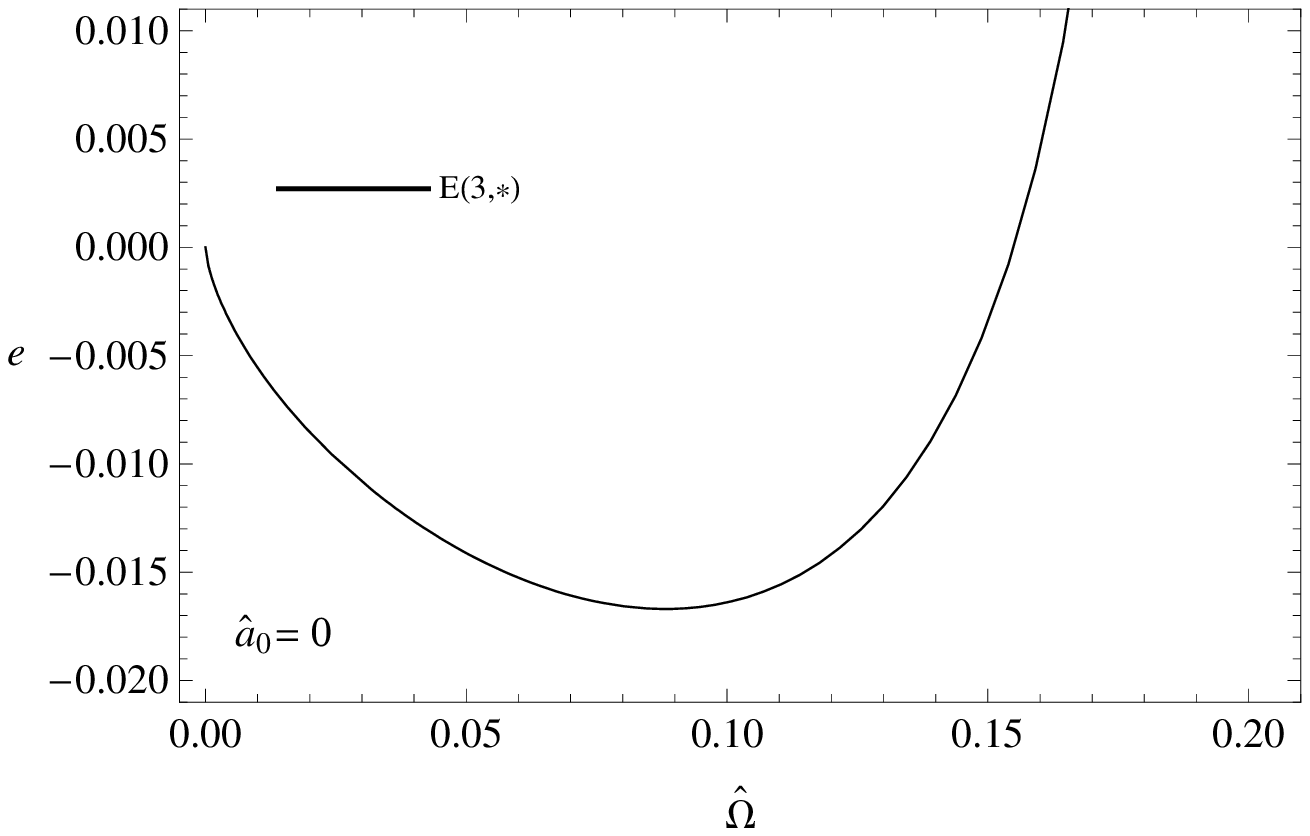}}&
\scalebox{0.65}{\includegraphics{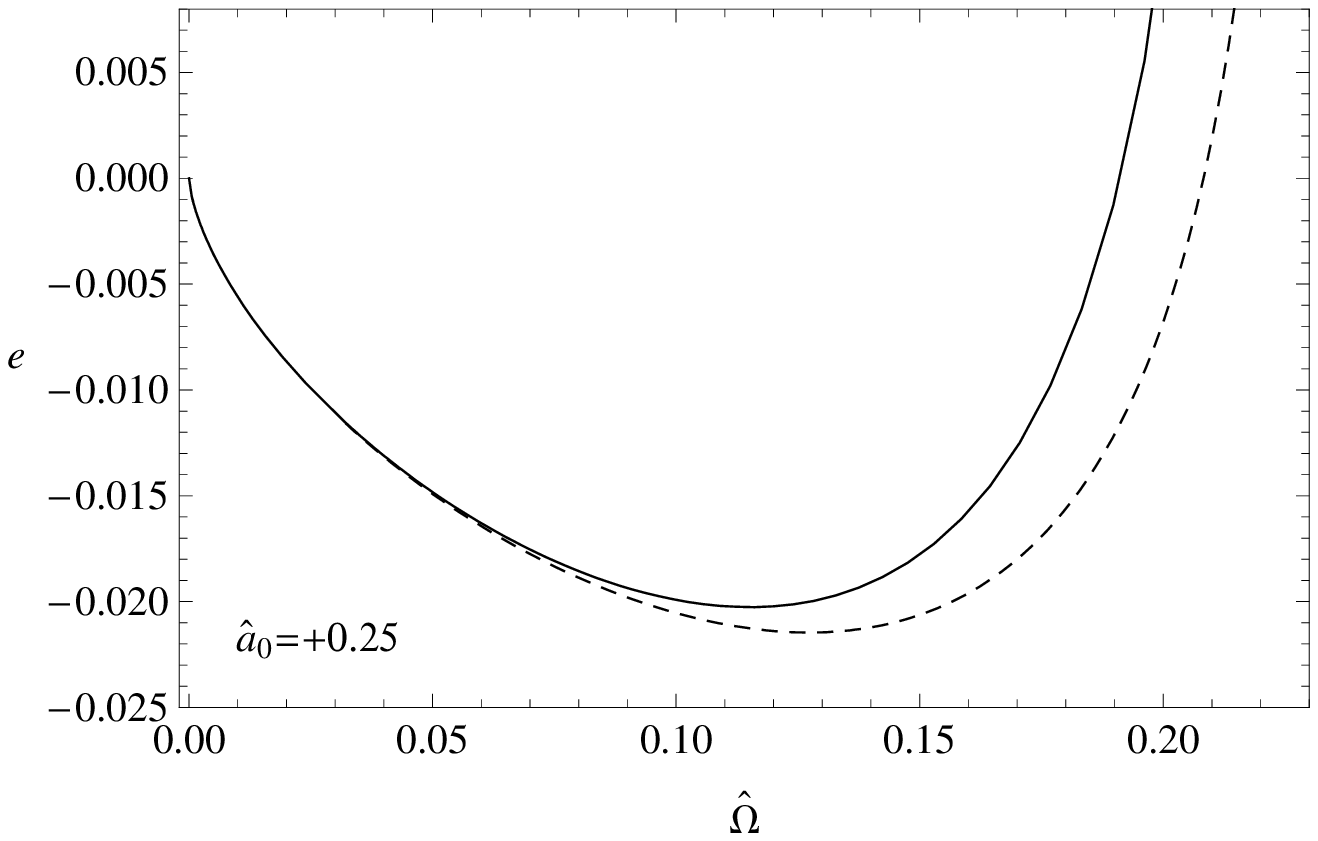}}\\[2ex]
\scalebox{0.65}{\includegraphics{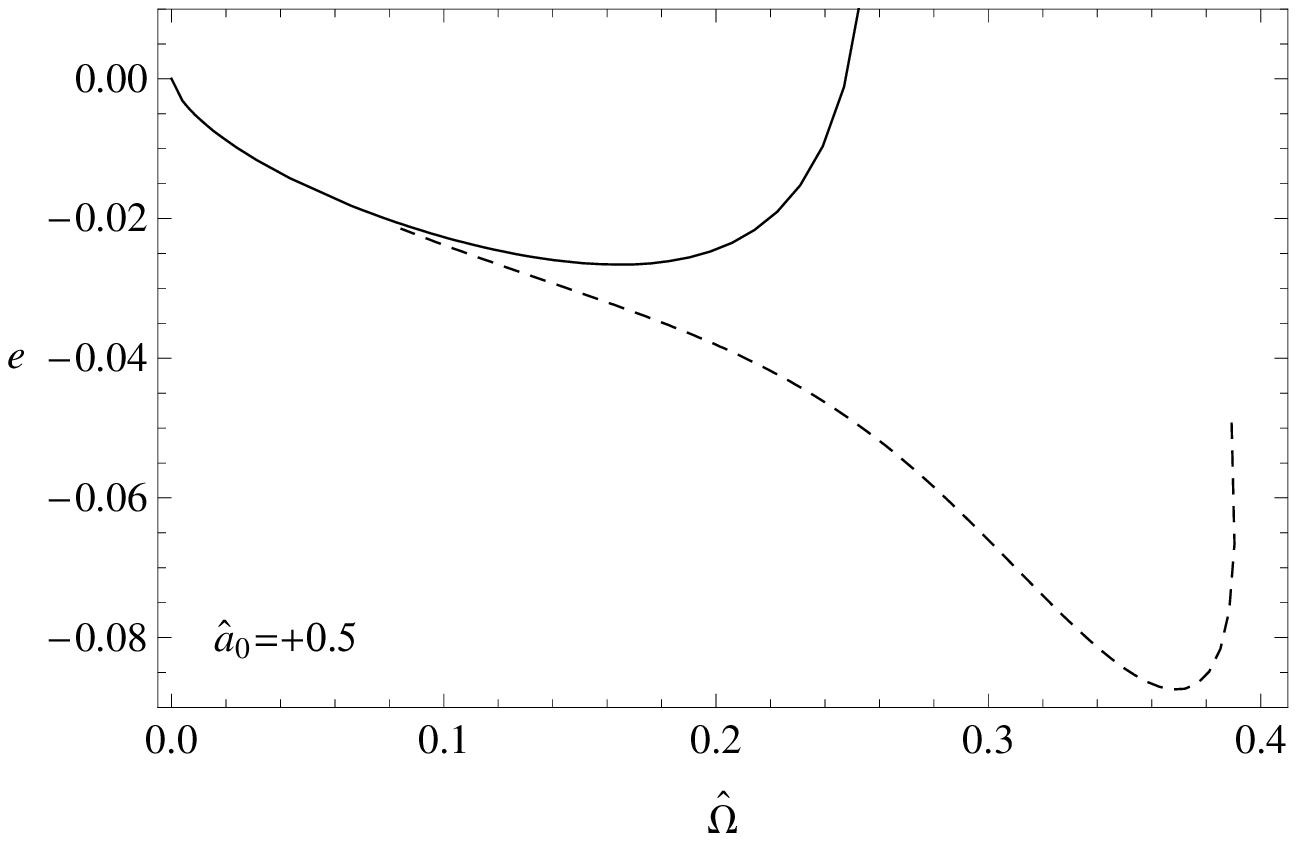}}&
\scalebox{0.65}{\includegraphics{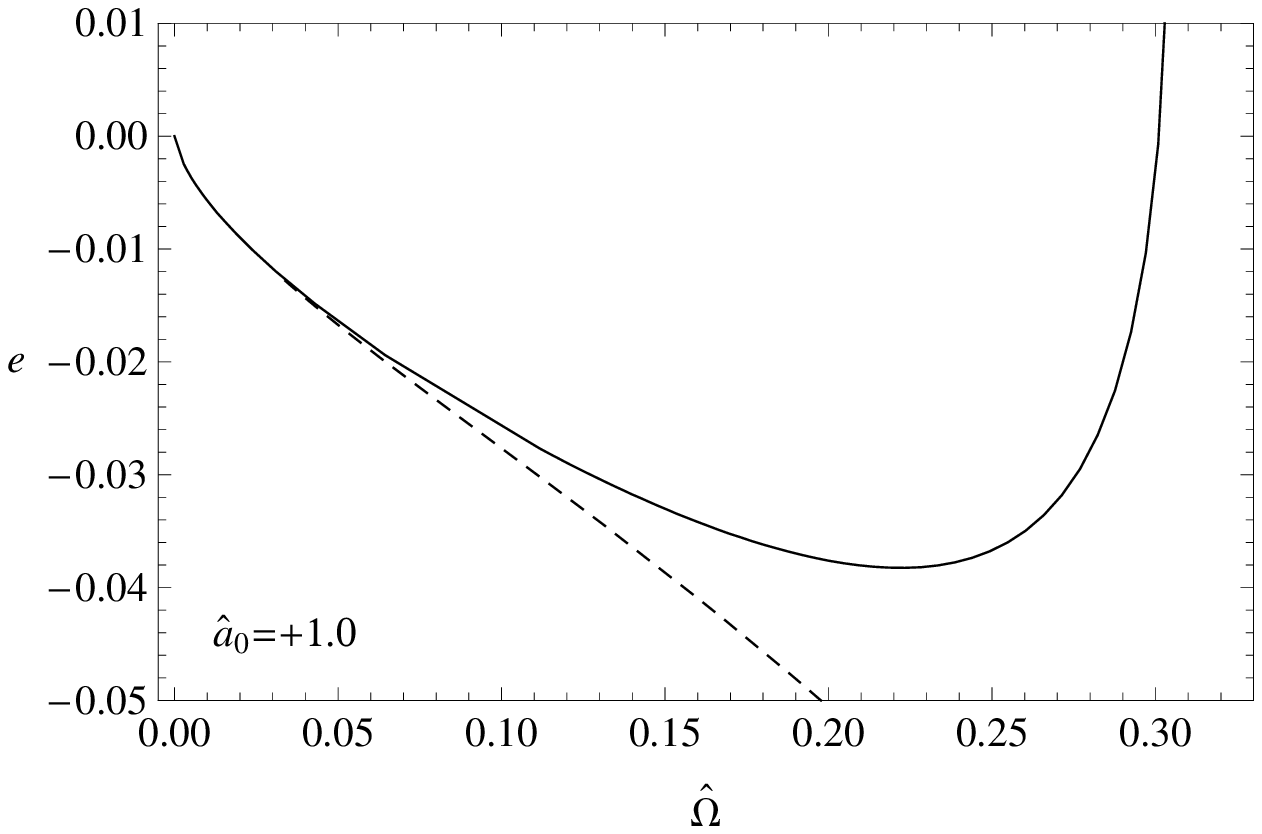}}
\end{tabular}
\caption{
Energy $e$ versus angular frequency $\hat{\Omega}$
along circular orbits for various values of the parameter $\hat{a}_0$,
as predicted by the EOB Hamiltonian.
We have assumed $m_1=m_2$, $a_1=a_2$, and $\theta_0=\pi/2$.
As before E$(n,s)$ refers to an EOB Hamiltonian, with $n$PN
accuracy in the orbital terms, and an accuracy in the spin-orbit coupling
equal to the LO one if $s=0$, and the NLO one if $s=1$. In all cases,
we include the full LO spin-spin coupling.}
\end{center}
\end{figure*}

Among the binding energy curves shown above, all the EOB ones
(at least when the effective spin is not too large and positive),
and some of the Taylor ones, admit a minimum for a certain
value of the orbital frequency $\hat{\Omega}$. This minimum corresponds
to an inflection point in the corresponding (EOB or Taylor) Hamiltonian
considered as a function of $r$. In other words, the minimum is
the solution of the two equations\footnote{Note in passing that,
in the EOB case, the two Eqs.\ \eqref{LSO} are equivalent to
the two similar equations involving the {\it effective} Hamiltonian:
$\pa H_\mathrm{eff}/\pa r=0$,
$\pa^2 H_\mathrm{eff}/\pa r^2=0$.}
\be
\label{LSO}
\frac{\pa H_\mathrm{real}}{\pa r} = 0, \quad
\frac{\pa^2 H_\mathrm{real}}{\pa r^2} = 0.
\ee
The solutions of these two simultaneous equations
correspond to what we shall call here
the Last Stable (circular) Orbit (LSO).\footnote{
As we recalled above, spinning binaries admit, in general, only
{\em spherical} orbits, rather than {\em circular} ones.
Reference \cite{Damour:2001tu} studied the binding energies
of the Last Stable Spherical Orbits (LSSO).
Here, however, we restrict ourselves to the parallel spin, where it makes
sense to study circular, equatorial orbits.} 
Several methods have been considered in the literature  
\cite{Damour:2000we,Blanchet:2001id}  for using PN-expanded
results to estimate the characteristics of the LSO. One of these methods
consists in considering the minima
{\it in the Taylor expansion of the function} $e(\hat{\Omega})$.
These minima (called ``Innermost Circular Orbit'' (ICO) in 
Refs. \cite{Blanchet:2001id,Blanchet:2006gy}, where they were used to estimate the
LSO of spinning binaries)
differ from the minima in the Taylor energy curves
considered in Fig.\ 1 above, which were based on using a Taylor-expanded
Hamiltonian. The advantage of consistently working (as we do here)
within a Hamiltonian formalism is that we are guaranteed that
the minima in the corresponding energy curves, when they exist,
do correspond to a Last Stable orbit
(and an associated inflection point) for some well-defined underlying dynamics.
By contrast the dynamical meaning (if any) of a minimum of the
Taylor-expanded function $e^{\rm Taylor} (\hat{\Omega})$ is unclear.
Anyway, as we saw above that the 3PN-accurate Taylor-expanded orbital
Hamiltonian does not admit any Last Stable Orbit, we have not plotted
in Fig.\ 4 the Taylor-based predictions for spinning binaries because
they do not seem to lead to reasonable results.

Concerning the dynamical meaning of the LSO,
let us  recall that it had been analytically
predicted in \cite{Buonanno:2000ef} (and confirmed in recent
numerical simulations \cite{Pretorius:2007nq}) that the transition
between inspiral and plunge is smooth and progressive, so that the
passage through the LSO is {\it blurred}. In spite of the
inherent ``fuzziness'' in the definition of the LSO, it is still
interesting to delineate its dynamical characteristics because they
strongly influence some of the gross features of the GW signal
emitted by coalescing binaries (such as the total emitted energy,
and the frequency of maximal emission).

\begin{figure}
\begin{tabular}{cc}
\scalebox{0.7}{\includegraphics{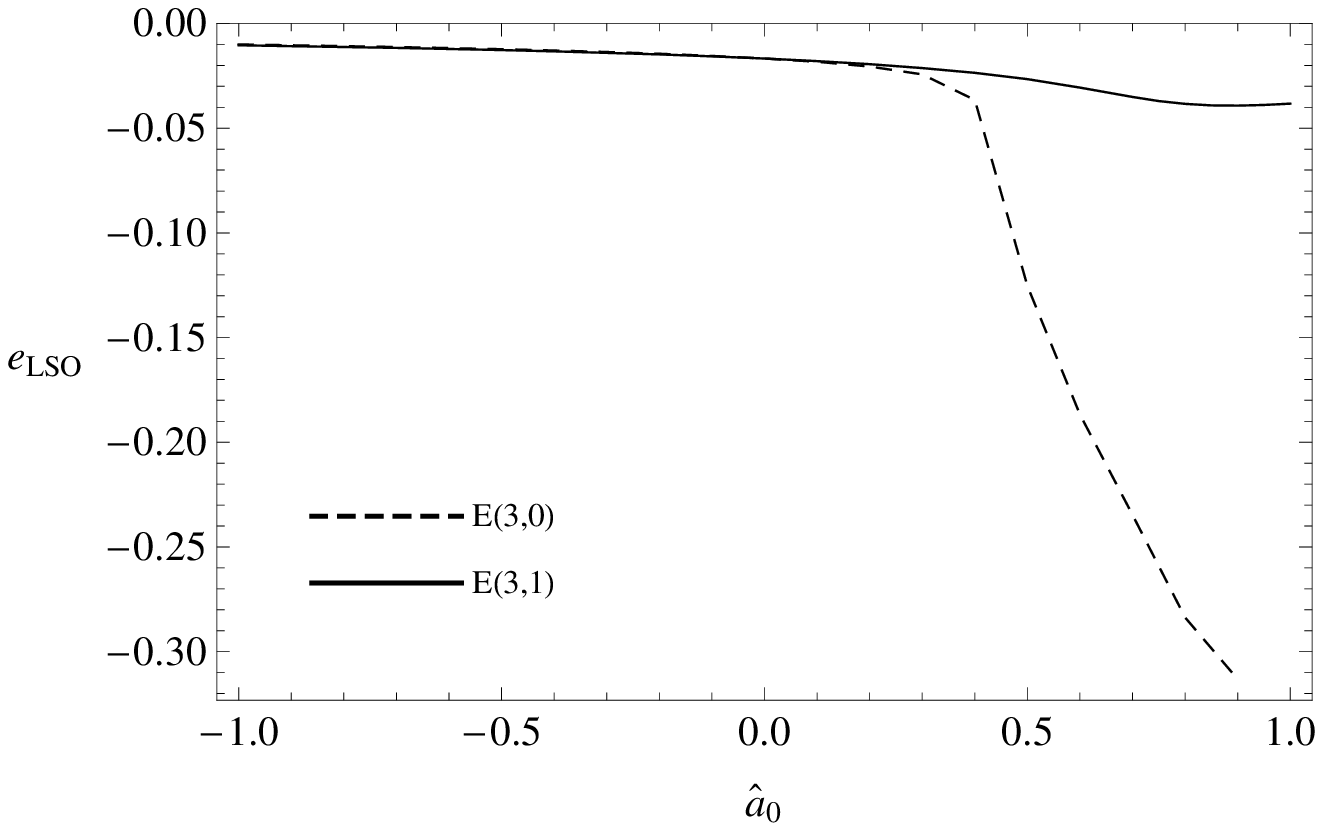}}\\
\scalebox{0.7}{\includegraphics{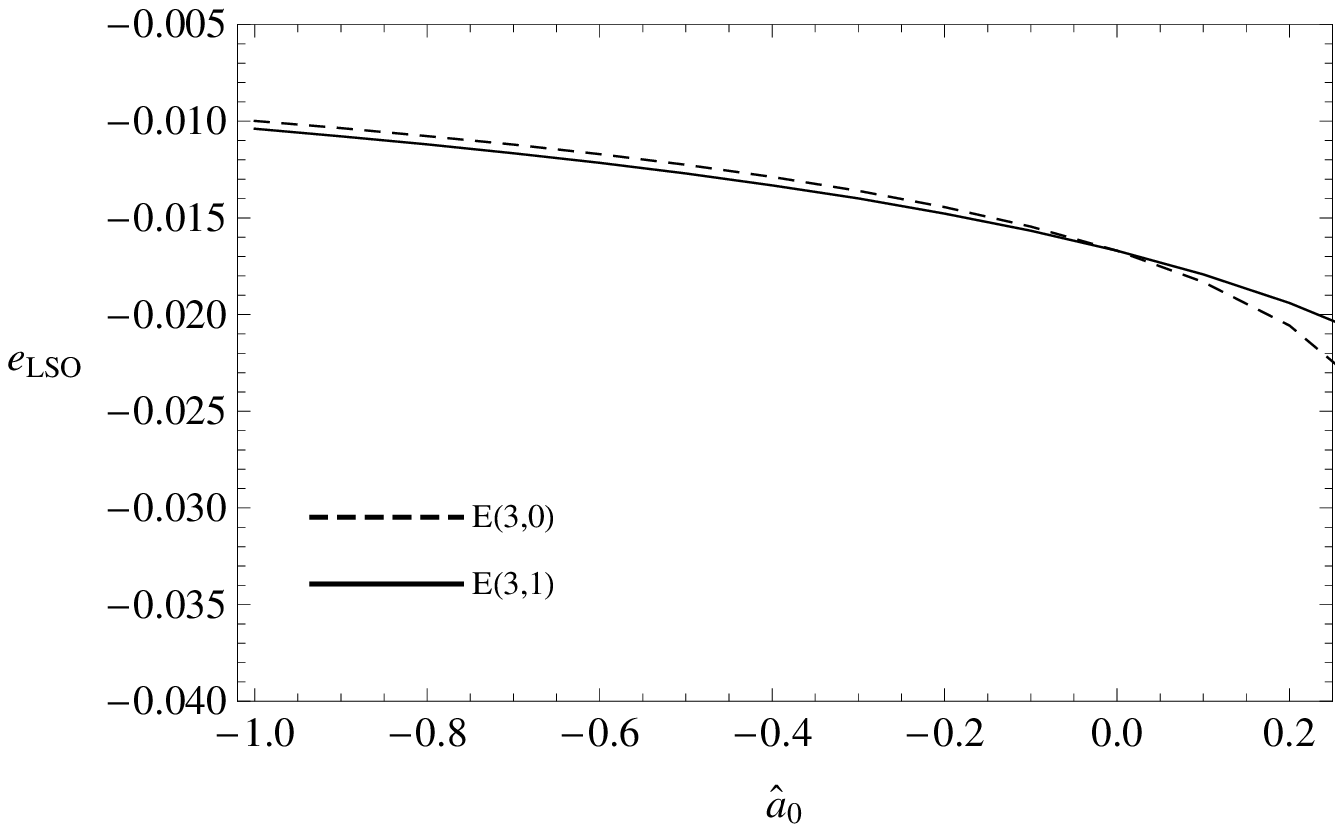}}\\
\scalebox{0.7}{\includegraphics{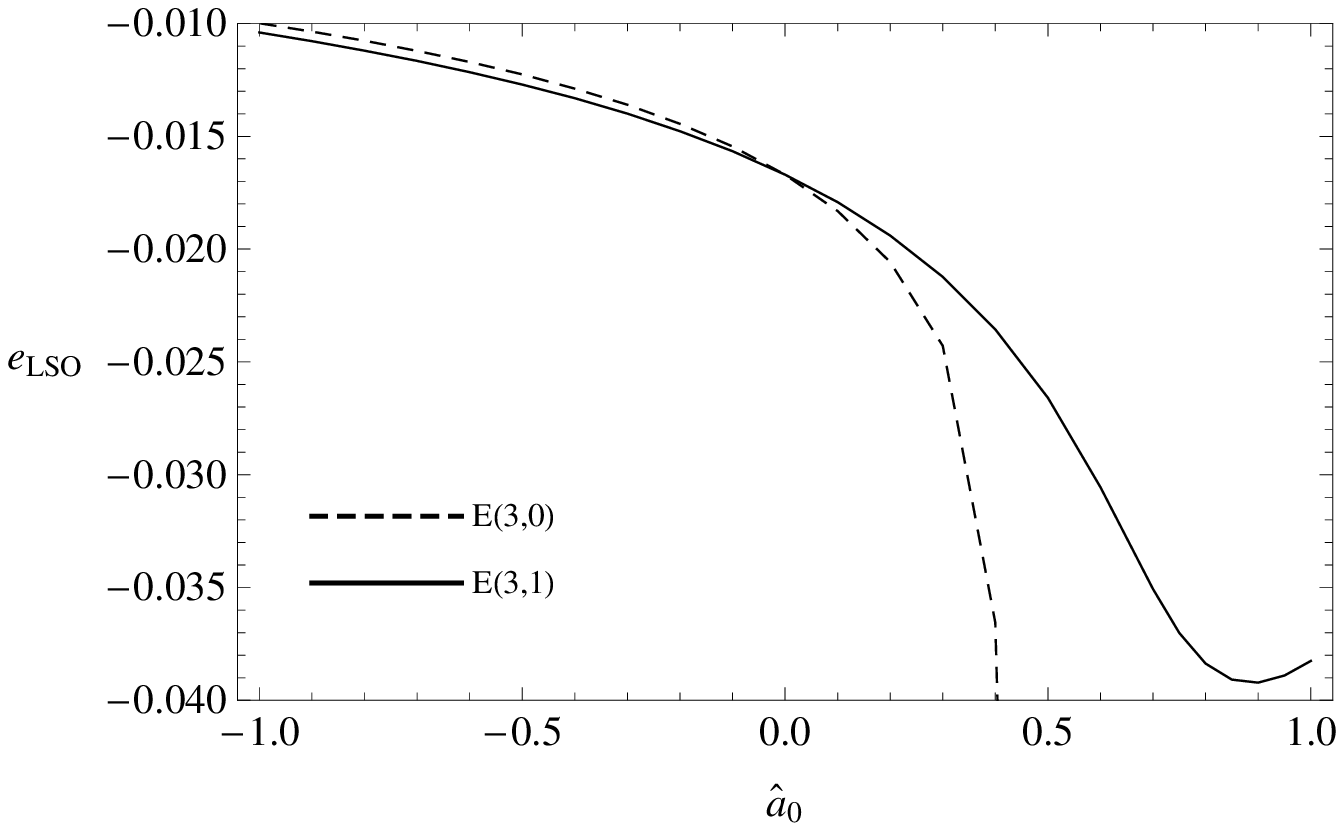}}
\end{tabular}
\caption{
Binding energy of the Last Stable (circular) Orbit (LSO)
predicted by the EOB approach. We study the effect of including
NLO spin-orbit terms by contrasting the LO and NLO predictions.
We plot the dimensionless energy $e_\mathrm{LSO}$ of the LSO
versus $\hat{a}_0$. We have assumed $m_1=m_2$, $a_1=a_2$,
and $\theta_0=\pi/2$. For E$(3,0)$ a LSO exists up to $\hat{a}_0\le+0.9$.}
\end{figure}

Let us comment on the results of our study
of the characteristics of LSO's:

\begin{itemize}
\item In Fig.\ 4 we plot the LSO binding energy, predicted by the
EOB approach, as a function of the
dimensionless effective spin parameter $\hat{a}_0$.
We contrast LO spin-orbit versus NLO spin-orbit (1 versus 0).
We use 3PN accuracy (for the orbital effects) in all cases, and always
include the LO spin-spin interaction.
The upper panel shows that the use of LO spin-orbit interactions
leads to dramatically negative LSO binding energies when the
spins become moderately large. [The middle panel is a close-up 
of the upper one, and focuses on spins $\hat{a}_0\leq+0.2$.]
We find that the 3PN-LO EOB Hamiltonian E$(3,0)$ admits
an LSO only up to spins as large as: $\hat{a}_0\leq+0.9$.
However, as first found in \cite{Damour:2001tu}, spin effects become
dramatically (and suspiciously) large already when $\hat{a}_0\geq+0.5$.
By contrast, as we found above, the inclusion of NLO spin-orbit interactions
has  the effect of {\it moderating} the dynamical influence of high (positive) spins.
The bottom panel focusses on our ``best bet''  3PN-NLO Hamiltonian E$(3,1)$.

As mentioned above, Ref.\ \cite{Blanchet:2006gy}
has considered, instead of the  Taylor-Hamiltonian LSO,
 the minimum of the Taylor-expanded function $e^{\rm Taylor}(\hat{\Omega})$ (or ``ICO'').
For the two cases $\hat{a}_0=-1,0$ (corresponding to their $\kappa_i=-1,0$),
they found, in the 3PN-NLO case, energy minima equal to
$e \equiv E_{\rm ICO}/m = -0.0116, -0.0193$
for corresponding orbital frequencies
$\hat{\Omega}\equiv m\omega_{\rm ICO}= 0.059, 0.129$.
These numerical values should be compared
with the numerical values we quote in Table I below.
On the other hand, for the large and parallel spin case $\hat{a}_0=+1$
Ref.\ \cite{Blanchet:2006gy} found that the Taylor-expanded
function $e^{\rm Taylor} (\hat{\Omega})$ has no minimum.
Finally, note that the qualitative shape of the curve giving the (EOB) LSO energy as a function
of the spin parameters $\hat{a}_0$
is similar both to the corresponding curve for a spinless test-particle in a Kerr background
(see, e.g., Fig.\ 7 below), and to the curve giving the LSO energy of a
{\it spinning test particle} in a Kerr background,
as a function of the test spin (see Fig.\ 4 in Ref.\ \cite{Suzuki:1997by}).
\end{itemize}

\begin{table}
\caption{
LSO parameters for {\it symmetric} binary systems
(with $m_1=m_2$ and $\hat{a}_1=\hat{a}_2=\hat{a}_0$)
for the 3PN-NLO EOB Hamiltonian E$(3,1)$.}
\begin{ruledtabular}
\begin{tabular}{ddd}
\hat{a}_0 & e & \hat{\Omega} \\ \hline
-1.00 & -0.01039 & 0.04473 \\
-0.75 & -0.01143 & 0.05139 \\
-0.50 & -0.01270 & 0.05989 \\
-0.25 & -0.01437 & 0.07143 \\
 0.00 & -0.01670 & 0.08822 \\
 0.25 & -0.02026 & 0.11521 \\
 0.50 & -0.02660 & 0.16444 \\
 0.75 & -0.03701 & 0.23249 \\
 1.00 & -0.03826 & 0.22210 \\
\end{tabular}
\end{ruledtabular}
\end{table}

To complement the information displayed in Figs.\ 1--4,
we give in Table I the numerical values of the main LSO characteristics
(binding energy and orbital frequency) for our ``best bet''
Hamiltonian, namely the 3PN-NLO EOB one E$(3,1)$.

In Figs.\ 5 and 6 we study the  effective-spin-dependence
of another LSO-related physical quantity of relevance for
the dynamics of coalescing binaries: the total (orbital plus spin)
angular momentum of the binary when it reaches the LSO
[i.e., at the end of the (approximately) adiabatic inspiral,
just before the plunge],
\be
{\bf J} \equiv {\bf L} + {\bf S}_1 + {\bf S}_2.
\ee
In terms of rescaled dimensionless variables, this becomes
\be
\hat{\mathbf{j}} \equiv \frac{c}{GM\mu} {\bf J}
= \hat{\bl} + \frac{m_1}{m_2} \hat{\bf a}_1 + \frac{m_2}{m_1} \hat{\bf a}_2,
\ee
where $\hat{\bl}\equiv c\,\bl$.
Actually, the most relevant quantity is the dimensionless Kerr parameter
associated to the total LSO mass-energy and the total LSO
angular momentum, i.e., the value at the LSO of the ratio
\begin{align}
\label{abh3}
\hat{a}_J \equiv \frac{cJ}{G\big(H_\mathrm{real}/{c^2}\big)^2}
= \nu \frac{\hat{j}}{\big(H_\mathrm{real}/{(Mc^2)}\big)^2},
\end{align}
where $\hat{j}$ is the modulus of $\hat{\bf j}$.

\begin{figure}
\scalebox{0.7}{\includegraphics{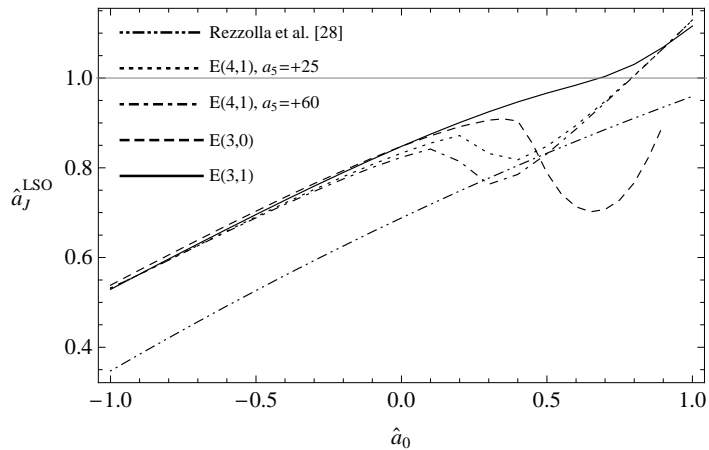}}
\caption{
The dimensionless total angular momentum Kerr parameter
$\hat{a}_J^{\mathrm{LSO}}$, Eq.\ \eqref{abh3}, at the LSO, versus $\hat{a}_0$.
We have assumed $m_1=m_2$, $a_1=a_2$, and $\theta_0=\pi/2$.
The parameter $\hat{a}_J^\mathrm{LSO}$ is computed from Eq.\ \eqref{abh3}
with $\hat{j}_\mathrm{LSO}=\hat{\ell}_\mathrm{LSO}+\hat{a}_1 + \hat{a}_2=
\hat{\ell}_\mathrm{LSO}+2\hat{a}_0$.
We compare the various EOB predictions obtained either by improving the
accuracy of spin-orbit terms [E$(3,1)$ versus E$(3,0)$], or by improving
the accuracy of orbital terms [E$(4,1)$ versus E$(3,1)$]. We use two
representative values of the 4PN parameter $a_5 = + 25$ and $a_5 = + 60$.
For comparison, we also include a fit to recent numerical estimates of
the {\it final} Kerr parameter of the black hole resulting from the
coalescence of the two constituent black holes.}
\end{figure}

\begin{figure}
\scalebox{0.7}{\includegraphics{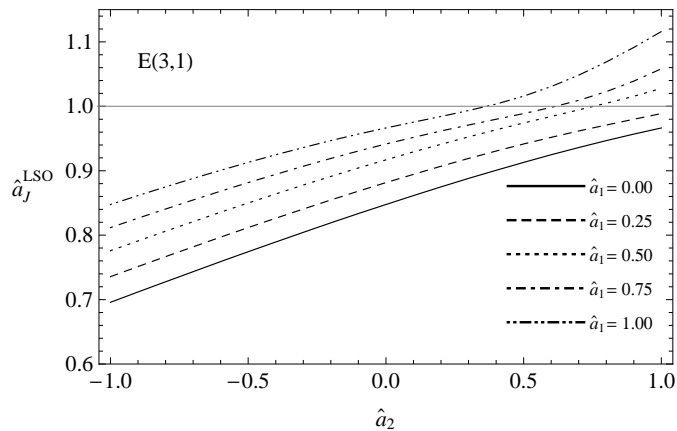}}
\caption{The dimensionless total angular momentum  Kerr parameter
$\hat{a}_J^{\mathrm{LSO}}$ at the E(3,1) LSO versus $\hat{a}_2$
for various values of the parameter $\hat{a}_1$. Here we consider
spin-dissymmetric systems with $a_1\neq a_2$
(but still $m_1=m_2$ and $\theta_0=\pi/2$).
The parameter $\hat{a}_J^\mathrm{LSO}$ is computed from Eq.\ \eqref{abh3}
with $\hat{j}_\mathrm{LSO}=\hat{\ell}_\mathrm{LSO}+\hat{a}_1+\hat{a}_2$.}
\end{figure}

\begin{figure}
\scalebox{0.7}{\includegraphics{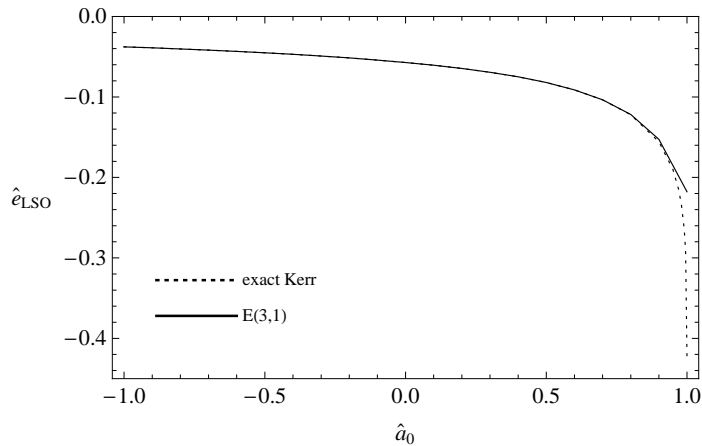}}
\caption{
Comparison of the test-mass limit $m_2\simeq\mu\to0$
(with fixed $\hat{a}_2$; so that $S_2/m_2\to0$)
for two Hamiltonians. We consider the specific ``non relativistic'' binding energy
$\hat{e}_{\rm LSO}\equiv e_{\rm LSO }/\nu = (E_{\rm LSO}-Mc^2)/(\mu c^2)$
at the LSO versus $\hat{a}_0$. The solid curve is the result of taking the test-mass limit of the
EOB Hamiltonian, while the short-dashed curve is the result for a test
particle moving in the Kerr  spacetime.}
\end{figure}

\begin{itemize}

\item In Fig.\ 5 we contrast the dependence of $\hat{a}_J^{\rm LSO}$
on the dimensionless effective spin parameter $\hat{a}_0$
for several EOB models: the two 3PN-accurate ones
[E$(3,0)$ using LO-accurate spin-orbit,
and E$(3,1)$ using NLO-accurate spin-orbit],
and two illustrative \cite{arXiv:0706.3732,arXiv:0712.3003}
``4PN-accurate'' NLO-spin-orbit models E$(4,1)$
(using either $a_5= +25$ or $a_5= +60$, as in Fig.\ 1).
[Here, we are still considering fully symmetric systems
with $m_1=m_2$ and $a_1=a_2$, so that $\hat{a}_0=\hat{a}_1=\hat{a}_2$.]
Again we see the moderating influence of NLO corrections.
The EOB-LO curve E$(3,0)$ exhibits a sudden drop down (pointed out
in \cite{Damour:2001tu}) before rising up again (and disappearing at
$\hat{a}_0 = +0.9$ when the LSO ceases to exist). By contrast,
the NLO curve E$(3,1)$ exhibits a much more regular dependence on
$\hat{a}_0$, which is roughly linear over the entire range
of values $-1\leq\hat{a}_0\leq1$. The two illustrative
E$(4,1)$ curves exhibit a ``mixed'' behaviour where a ``drop'' similar
to the one featuring in  the
LO curve is still present, though it is moderated by  NLO spin-orbit effects.
This sensitivity to the inclusion of a 4PN contribution in
$A(\hat{u})$ is due to a delicate interplay between the modified shape
of the basic spin-independent ``radial potential''
$A(\hat{u},a_5)$ and the use of a (1,4) Pad\'e resummation of
the ``effective spin-dependent radial potential''
$\Delta_t(R)$, Eq. \eqref{deltat}. Indeed, the additional
contributions proportional to $a_5$ and $a^2$ are {\it both
repulsive }, and tend to compound their effect, which is to
push the LSO toward closer, more bound orbits \cite{Damour:2001tu}.

We have also indicated in Fig.\ 5 the {\em final}
(i.e., after coalescence) dimensionless Kerr parameter
of (symmetric) spinning binaries, as obtained in recent numerical simulations
\cite{Herrmann:2007ex,Marronetti:2007wz,Rezzolla:2007xa,Rezzolla:2007rd}.
For simplicity, we have shown the simple analytic fit proposed in \cite{Rezzolla:2007xa}.
The fact that the 3PN-NLO-accurate EOB LSO Kerr parameter [E$(3,1)$]
is systematically {\it above} the final Kerr parameter
is in good agreement with the fact that, after reaching the LSO,
the system will still loose
a significant amount of angular momentum\footnote{
We use here the fact (found in numerical calculations, and implied by the analytical EOB approach),
that, fractionally speaking,
the angular momentum loss after the LSO is significantly higher than the
corresponding energy loss.}
during the plunge and the merger-plus-ringdown.
In the case of {\em non-spinning} binaries,
it has been shown that, by using the EOB formalism up to the end
of the process [i.e., by taking into account the losses of $J$ and $E$
during plunge, as well as during merger-plus-ringdown], there
was a good agreement (better than $\sim2\%$) between EOB and numerical relativity
for the final spin parameter \cite{Damour:2007cb}.
We hope that the same type of agreement will hold also
in the case of {\it spinning} binaries considered here.

\item In Fig.\ 6 we plot the LSO dimensionless Kerr parameter of Eq.\ \eqref{abh3} for
{\it spin-dissymmetric} systems, namely $a_1 \neq a_2$ (but with $m_1=m_2$), computed
with the 3PN-NLO EOB Hamiltonian model E$(3,1)$. This plot illustrates
that the LSO spin parameter is a smooth (and essentially linear)
function of the two individual spins.

\item Finally, we compare in Fig.\ 7 the spinless test particle limit
[i.e., $m_2 \to 0$, together with $a_2 = S_2/(m_2c) \to 0$,
as appropriate to black holes for which $\hat{a} \leq 1$]
for two Hamiltonians: the 3PN-NLO EOB one E$(3,1)$,
and the exact one, as known from the geodesic action of a spinless test particle in the Kerr metric.
For non-spinning systems the EOB Hamiltonian is constructed so as to reduce to the exact
Schwarzschild-derived one in the test-particle limit.
However, for spinning systems, we have chosen in Eq.\ \eqref{deltat}
to define the crucial metric coefficient $\Delta_t(R)$
by Pad\'e-resumming the sum of $A(R;\nu)+a^2/R^2$.
This Pad\'e-resummation is indeed useful for generally ensuring,
for comparable mass systems, that $\Delta_t(R)$ have a simple zero
at some ``effective horizon'' $r_\mathrm{H}$.
However, in the test-mass limit $\nu\to0$,
while the Taylor-approximant to $A(R;\nu)+a^2/R^2$ would coincide with the exact Kerr answer,
the Pad\'e-resummed version of $A(R;\nu)+a^2/R^2$ differs from it. We see, however, on Fig.\ 7
that the resulting difference has a very small effect on the LSO energy per unit ($\mu$) mass,
except when the dimensionless effective spin $\hat{a}_0$ is very close to $+1$.
On the other hand, as we saw above when discussing Fig.\ 5, the
issue of the Pad\'e resummation of $\Delta_t(R)$ becomes more subtle
when one considers the comparable-mass case, together with the
inclusion of a repulsive 4PN parameter $a_5$.

\end{itemize}

\section{Conclusions}

The main conclusions of this work are:

\begin{itemize}

\item We have prepared the ground for an accurate Effective One Body (EOB) description
of the dynamics of binary systems made of {\it spinning} black holes by incorporating
the recent computation of the next-to-leading order (NLO) spin-orbit interaction Hamiltonian 
\cite{Damour:2007nc} (see also Refs.\ \cite{Faye:2006gx,Blanchet:2006gy}) into a previously
developed extension of the EOB approach to spinning bodies \cite{Damour:2001tu}.

\item We found that the inclusion of NLO spin-coupling terms has the quite significant result
of {\it moderating} the effect of the  LO spin-coupling, which would, by itself (as found
in Ref.\ \cite{Damour:2001tu}), predict that the Last Stable (circular) Orbit (LSO)
of  parallely-fast-spinning black holes can reach very large binding energies of the
order of $30\%$ of the total rest-mass energy $Mc^2$. By contrast, the inclusion of
NLO spin-orbit terms predicts that the LSO of parallely-fast-spinning systems, though
significantly more bound than that of non-spinning holes, can only reach binding energies
of the order of $4\%$ of the total rest-mass energy $Mc^2$ (see Fig.\ 4 above). This reduction
in the influence of the spin-orbit coupling is due to the fact that the (effective) 
``gyro-gravitomagnetic ratios'' are {\it reduced} by NLO effects from their LO values
$g^\mathrm{LO}_S=2$, $g^\mathrm{LO}_{S^*}=\frac{3}{2}$ to the values (here considered
along circular orbits)
\begin{align}
g^\mathrm{circ\,eff}_S &= 2 - \frac{5}{8} \nu x,
\nonumber\\[1ex]
g^\mathrm{circ\,eff}_{S^*} &= \frac{3}{2} - \Big(\frac{9}{8}+\frac{3}{4}\nu\Big) x,
\end{align}
where $x\simeq GM/(R c^2)\simeq(GM\Omega/c^3)^{2/3}$. This reduction then reduces the
{\it repulsive} effect of the spin-orbit coupling which is responsible for allowing
the binary system to orbit on very close, and very bound, orbits
(see discussion in Section 3C of Ref.\ \cite{Damour:2001tu}).

\item We studied the dependence of the dimensionless Kerr parameter of the binary system,
$\hat{a}_J\equiv{cJ}/\boldsymbol(G(H_\mathrm{real}/{c^2})^2\boldsymbol)$, computed
at the LSO, on the spins of the constituent black holes. Again the moderating effect
of including NLO spin-orbit terms is very significant
(compare the solid and the dashed\footnote{
Compare also with Fig.\ 2 of Ref.\ \cite{Damour:2001tu}
where the relevant LO result is the curve labelled ``DJS''
which reaches a  maximum around
$\hat{a}\equiv\frac{7}{8}\hat{a}_0\simeq0.31$,
in agreement with the (local) maximum in the dashed
line of our Fig.\ 5 reached around $\hat{a}_0 \simeq 0.36$.}
lines in Fig.\ 5). Thanks to this moderating effect the LSO Kerr parameter
$\hat{a}_J^{\rm LSO}$ is found to have a monotonic, and roughly linear, dependence
on the spin parameters of the individual black holes
(see solid line in Fig.\ 5 and the various curves in Fig.\ 6). We also studied the effect of
including the type of 4PN parameter $a_5$ found useful in recent work
\cite{arXiv:0706.3732,arXiv:0711.2628,arXiv:0712.3003,DamourNagar08}
for improving the agreement between EOB waveforms and numerical ones.

\item We leave to future work the analog of what was initiated
for spinning systems in Ref.\ \cite{Buonanno:2005xu},
and recently completed for the case of non-spinning black holes
in Ref.\ \cite{Damour:2007cb}, i.e., a full dynamical study,
within the EOB approach, of the Kerr parameter of the {\it final} black hole
resulting from the merger of spinning black holes
which takes into account the angular momentum losses that occur after the LSO,
during the plunge, the merger, and the ringdown.
Let us also note that Ref.\ \cite{Buonanno:2007sv} has recently proposed 
an approximate analytical approach (which is similar in spirit to the
approximation used in Refs.\ \cite{Buonanno:2000ef,Damour:2001tu,Buonanno:2005xu}
and above, namely that of considering the Kerr parameter of an effective test particle
at, or after, the LSO) towards estimating the final spin of a binary black hole coalescence.
The resulting prediction is, however, only in coarse agreement $\sim10\%$
with numerical results. Note in this respect that,
as displayed in Fig.\ 5, the ``zeroth order'' EOB result
[corresponding to using the  Kerr parameter for E$(3,1)$ at the LSO,
without taking into account the later losses of angular momentum] is already
in  $\sim20\%$  agreement with the fit to the numerical data \cite{Rezzolla:2007xa}.
The fact (displayed on Fig.\ 5) that the E(3,1)
EOB LSO Kerr parameter is systematically {\it above} the final
(after coalescence) Kerr parameter determined by recent numerical simulations
\cite{Herrmann:2007ex,Marronetti:2007wz,Rezzolla:2007xa,Rezzolla:2007rd}
is in qualitative agreement with the fact that the system will loose a significant
amount of angular momentum during the plunge and the merger-plus-ringdown.
Note, however, the sensitivity of $\hat{a}_J^{\rm LSO}$ to a
``4PN deformation'' of the EOB Hamiltonian by the parameter $a_5$.
As said above, this sensitivity is due to the fact that
the radial function $\Delta_t(R)/R^2$ combines the additional repulsive
effects of both a positive 4PN contribution
$+a_5\nu\boldsymbol(GM/(c^2R)\boldsymbol)^5$
and a positive spin-dependent contribution
$+a^2/R^2$. We leave to future work an exploration of this issue,
which might need the use of a different Pad\'e resummation than
the (1,4) one used in \eqref{deltat}.

It remains to be seen whether the EOB/Numerical Relativity comparison
for the final Kerr parameter of spinning systems will be as good as it was found to be
for the non-spinning case \cite{Damour:2007cb}, i.e., at the $2\%$ level.
If this is the case, it will establish the physical relevance
of the improved EOB Hamiltonian constructed in the present paper.

\item Let us finally note that there is some {\it flexibility} in the improved spin-dependent EOB
Hamiltonian proposed above (besides the flexibility in the choice
of the Pad\'e resummation mentioned above).
On the one hand, the choice \eqref{ab} for the
gauge parameters $a(\nu)$ and  $b(\nu)$ might be replaced by other choices.
On the other hand, the choice \eqref{defseff} for the effective spin vector might
also be replaced by other ones. In particular, it might be interesting to consider
the alternative definition
\begin{align}
\label{defseffnew}
Mc\,\mathbf{a}_{\rm new} &\equiv \mathbf{S}_\mathrm{eff\,new}
\equiv \frac{1}{2} g^\mathrm{eff\,new}_S \mathbf{S}_0
\nonumber\\
&= \frac{1}{2} g^\mathrm{eff\,new}_S \big( {\bf S} + {\bf S}^* \big).
\end{align}
This definition coincides with the one used above at LO in spin-orbit effects 
(because $g^\mathrm{eff\,new}_S = 2 + \mathcal O (\nu/c^2)$), and allows one to use
a simplified supplementary spin-orbit contribution, built with
\be
\label{sigmanew}
\bsigma^{\rm new} \equiv \frac{1}{2}
\big(g^\mathrm{eff}_{S^*}-g^\mathrm{eff}_S\big)\mathbf{S}^*,
\ee
instead of \eqref{sigma}. It might be interesting to explore which of these
possible definitions exhibits the best agreement with current numerical results.

\end{itemize}

\begin{acknowledgments}
This work was supported in part
by the KBN Grant no 1 P03B 029 27 (to P.J.)
and by the Deutsche Forschungsgemeinschaft (DFG)
through SFB/TR7 ``Gravitational Wave Astronomy''.
\end{acknowledgments}

\end{document}